\documentclass[a4paper,12pt]{article}
\pdfoutput=1 

\usepackage{jheppub} 

\usepackage{tabularx,subcaption,fancyhdr,mathtools,physics}
\usepackage[makeroom]{cancel}
\usepackage{bm}
\usepackage{csquotes}
\usepackage[nottoc,notlot,notlof]{tocbibind}
\usepackage{float}
\usepackage{relsize}
\usepackage{hyperref}
\usepackage{color,latexsym, array,multirow,url,verbatim, enumerate,cancel}
\DeclareUnicodeCharacter{2212}{-}
\hypersetup{colorlinks=true,linkcolor=blue,citecolor=magenta,filecolor=magenta,
	urlcolor=cyan}
\def \gl{\widetilde g}
\def \st1{\widetilde t_1}
\def \mst1{m_{\st1}}
\def \sbot1{\widetilde b_1}

\def \lspone{\widetilde\chi_1^0}
\def \mlspone{m_{\lspone}}
\def \lsptwo{\widetilde\chi_2^0}
\def \mlsptwo{m_{\lsptwo}}

\def\chonepm{\widetilde{\chi}_1^{\pm}}
\def\chonemp{\widetilde{\chi}_1^{\mp}}

\def\mchonepm{m_{\chonepm}}

\newcommand{\beq}{\begin{equation}}
\newcommand{\eeq}{\end{equation}}
\def\bea{\begin{eqnarray}}
\def\eea{\end{eqnarray}}

\def \met{\rm E{\!\!\!/}_T}

\title{Markov Chain Monte Carlo analysis to probe trilinear $R$-parity violating SUSY scenarios and possible LHC signatures }

\author[a]{Arghya Choudhury,} 
\author[b]{Sourav Mitra,}
\author[a]{Arpita Mondal,}
\author[c]{and Subhadeep Mondal}

\affiliation[a]{Department of Physics, Indian Institute of Technology Patna, Bihar, 801106, India}
\affiliation[b]{Surendranath College, 24/2 M. G. Road, Kolkata, West Bengal, 700009, India}
\affiliation[c]{Department of Physics, SEAS, Bennett University, Greater Noida, Uttar Pradesh, 201310, India}

\emailAdd{arghya@iitp.ac.in}
\emailAdd{hisourav@gmail.com}
\emailAdd{arpita\_1921ph15@iitp.ac.in}
\emailAdd{subhadeep.mondal@bennett.edu.in}

\abstract{In this article, we probe the trilinear $R$-parity violating (RPV) supersymmetric (SUSY) scenarios with specific nonzero interactions in the light of neutrino oscillation, Higgs, and flavor observables. We attempt to fit the set of observables using a state-of-the-art Markov Chain Monte Carlo (MCMC) setup and study its impact on the model parameter space. Our main objective is to constrain the trilinear couplings individually, along with some other SUSY parameters relevant to the observables. We present the constrained parameter regions in the form of marginalized posterior distributions on different two-dimensional parameter planes. We perform our analyses with two different scenarios characterized by our choices for the lightest SUSY particle (LSP), bino, and stop. Our results indicate that the lepton number violating trilinear couplings $\lambda_{i33}$ ($i$=1,2) and $\lambda_{j33}^{\prime}$ ($j$=1,2,3) can be at most of the order of $10^{-4}$ or even smaller while $\tan\beta$ is restricted to below 15 even when $3\sigma$ allowed regions are considered. We further comment on the possible LHC signatures of these LSPs focusing on and around the best-fit regions.}

\begin{document} 
\maketitle

\section{Introduction}
\label{sec:intro}
Neutrino oscillation data from several experiments \cite{Borexino:2013zhu,KamLAND:2013rgu,RENO:2018dro,DayaBay:2018yms,Super-Kamiokande:2019gzr,T2K:2018rhz,NOvA:2019cyt,deSalas:2020pgw} provides some of the strongest evidence for the existence of beyond the standard model (BSM) physics. The mass-squared differences and mixing angles measured by neutrino oscillation experiments indicate that at least two light neutrino masses are nonzero, along with significant mixing among the three flavor states.
Standard model (SM) \cite{GLASHOW1961579,PhysRevLett.19.1264,Salam:1968rm,GELLMANN1964214} as well as $R$-Parity \cite{Martin:1997ns} conserving minimal supersymmetric (SUSY) standard model (MSSM) \cite{Drees:2004jm, Baer:2006rs, Martin:1997ns} cannot explain the generation of light neutrino masses and mixings. However, $R$-Parity happens to be a discrete symmetry that is imposed on MSSM to ensure the stability of the lightest supersymmetric particle (LSP) and does not follow from any fundamental principle. A more generic version of the MSSM, therefore, contains $R$-parity violating (RPV) terms \cite{Martin:1997ns,Barbier:2004ez,Choudhury:2024ggy,Allanach:2003eb}, which can lead to generation of light neutrino masses and mixings naturally \cite{Grossman:1997is,Grossman:2003gq,Rakshit:2004rj,Barbier:2004ez, Hempfling:1995wj, Allanach:1999ic}. 

Unlike $R$-parity conserving scenarios, RPV scenarios are yet to be studied exhaustively. It would be of particular interest to see how much of its sub-TeV parameter space can be constrained from existing data from neutrino oscillation as well as precision Higgs data and collider limits on SUSY particles. Four types of RPV terms can be added to the superpotential, which either violate the lepton number or the baryon number. Among these, there are three terms, one bilinear and two trilinear, that violate the lepton number by one unit, and one trilinear term, which violates the baryon number by one unit. The implication of these nonzero RPV couplings on light neutrino masses and mixings have been discussed in existing literature \cite{Grossman:1997is,Grossman:2003gq,Rakshit:2004rj,Barbier:2004ez,PhysRevD.78.075021,Hundi:2011si,Diaz:2014jta, Allanach:1999ic,Hempfling:1995wj, deCampos:2012pf, Hirsch:2000ef}. However, a detailed statistical analysis through posterior distributions of the RPV couplings and other relevant SUSY parameters was only performed recently with nonzero bilinear RPV couplings \cite{Choudhury:2023lbp}. A similar study with trilinear lepton number violating couplings has been missing so far. Note that, whereas the bilinear RPV couplings can partially generate light neutrino masses at the tree level itself, the lowest order contributions from the trilinear couplings arise at one loop \cite{Grossman:1997is,Rakshit:2004rj,Grossman:2003gq,Barbier:2004ez, Hempfling:1995wj}. Nevertheless, a complete one loop calculation is necessary for both scenarios. 

In this work, we focus on chosen trilinear lepton number violating couplings that contribute dominantly towards light neutrino mass generation. Apart from these, we set all other RPV couplings to zero. Keeping both bilinear and trilinear lepton number violating interactions makes it easier to fit the neutrino oscillation data, with the dominating contribution arising from the bilinear terms. However, our objective here is to assess the impact of the neutrino oscillation data on lepton number violating trilinear couplings only. Note that just as one can start only with the bilinear interaction terms and generate the lepton number violating trilinear interactions \cite{Roy:1996bua}, starting from the trilinear interaction terms through renormalization group evolution, one can also generate the bilinear terms at a different energy scale \cite{Nardi:1996iy,deCarlos:1996ecd}. In addition to the RPV couplings and their corresponding soft terms, some other SUSY parameters are also relevant in this context. These include the $\mu$ parameter, $\tan\beta$, and the soft mass parameters of sleptons and squarks.        
Choice of these SUSY parameters can, in turn, affect the 125 GeV Higgs observables \cite{ATLAS:2021vrm,cms_web1}, which are also taken into account. In addition to that, $B$-meson decay branching ratios \cite{Archilli:2017xmu,HFLAV:2019otj,LHCb:2021vsc} that are very precisely measured and are sensitive to choice of SUSY spectrum are also considered in this study as observables. 
We have performed our analyses for two different LSP scenarios, namely, bino and stop. First, we consider the model with only LQD type couplings (\texttt{Bino$_{\lambda^\prime}$-model}). Next, we analyze the models with both LLE and LQD couplings  (\texttt{Bino$_{\lambda \lambda^\prime}$-model}). Finally, we consider the stop LSP scenario, with both LLE and LQD couplings (\texttt{Stop$_{\lambda \lambda^\prime}$-model}).
We also comment on the impact of collider limits on SUSY particle masses in RPV context \cite{ATLAS:2021yyr,Barman:2020azo,ATLAS:2021moa,ATLAS:2023lfr, ATLAS:2017jvy, ATLAS:2024zkx, CMS:2020bfa, ATLAS:2019lff} on our results.

Given the precision of the observables used in this study and the number of input parameters, a random scan does not yield any allowed parameter space. Hence, we have adopted a Markov Chain Monte Carlo (MCMC) \cite{foreman2013emcee} algorithm to sample the parameter space. The objective is not only to find some allowed parameter space but to generate posterior distributions of the input parameters corresponding to $99\%$, $95\%$, and $68\%$ confidence levels through Bayesian analysis similar to the study performed recently with only bilinear interaction terms \cite{Choudhury:2023lbp}. The derived results not only give us a robust understanding of the available parameter space but also provide future directions to any phenomenological studies performed with this model.

The paper is organized as follows: 
In Sec.~\ref{sec:model}, we discuss the generation of neutrino masses and mixing from the trilinear lepton number violating terms. We summarize the 
latest global fit of the neutrino oscillation data along with 
the constraints coming from Higgs mass, Higgs coupling strength modifiers, and the $B$-meson decay branching ratios in Sec.~\ref{sec:observable}. 
In Sec.~\ref{sec:constraints_mass}, we review the recent collider limit on the sparticle masses coming from direct search results obtained by the ATLAS and CMS groups. The details of MCMC analysis setup are briefly discussed in Sec.~\ref{sec:analysis_details}. In Sec.~\ref{sec:model_bino} and Sec.~\ref{sec:squark_lsp}, we present the results for different trilinear 
RPV models considering bino and stop LSP scenarios, respectively.
Lastly, we conclude in Sec.~\ref{sec:conclusion}.

\section{Model description}
\label{sec:model}
The lepton number violating terms associated with trilinear couplings in the RPV SUSY model are expressed as follows~\cite{Barbier:2004ez, Dreiner:1997uz, Banks:1995by}
\begin{equation}
\label{eq:potential}
W_{\cancel{L}} = \frac{1}{2}\lambda_{ijk}L_iL_jE_k^c + \lambda^{\prime}_{ijk}L_iQ_jD_k^c
\end{equation}
The first term can be reformulated using $SU(2)_L$ indices as $\epsilon_{\alpha\beta}L_i^{\alpha}L_j^{\beta}E_k^c$, where $\alpha, \beta = 1, 2$. Due to the gauge invariance and presence of antisymmetric $\epsilon_{\alpha\beta}$ tensor, the $\lambda_{ijk}$ couplings also become antisymmetric in the first two indices ($\lambda_{ijk} = -\lambda_{jik}$)~\cite{Barbier:2004ez} and the first term in the Eq.~\ref{eq:potential} can be written as $\lambda_{ijk}\hat{L}_i\hat{L}_j\hat{E}_k^c = \lambda_{ijk}\epsilon_{\alpha\beta}\hat{L}^{\alpha}_i\hat{L}^{\beta}_j\hat{E}_k^c = -\lambda_{jik}\epsilon_{\beta\alpha}\hat{L}^{\alpha}_j\hat{L}^{\beta}_i\hat{E}_k^c = -\lambda_{jik}\hat{L}_i\hat{L}_j\hat{E}_k^c$. This leads to 9 unique $\lambda_{ijk}$  and 27 $\lambda_{ijk}^{\prime}$ lepton number violating trilinear couplings. The Lagrangian for these two terms can be represented as~\cite{Dreiner:1997uz, Barbier:2004ez}
\begin{equation}
\label{eq:lag}
\begin{split}
\mathcal{L}_{\cancel{L}} & = -\frac{\lambda_{ijk}}{2}[\tilde{\nu}_{iL}\bar{l}_{kR}l_{jL} 
 + \tilde{l}_{jL}\bar{l}_{kR}\nu_{iL} + \tilde{l}_{kR}^*\bar{\nu}_{iR}^cl_{jL} - (i\leftrightarrow j)] \\
&  -\lambda_{ijk}^{\prime}[\tilde{\nu}_{iL}\bar{d}_{kR}d_{jL} + \tilde{d}_{jL}\bar{d}_{kR}\nu_{iL} + \tilde{d}_{kR}^*\bar{\nu}_{iR}^cd_{jL} - \tilde{l}_{iL}\bar{d}_{kR}u_{jL} \\
& - \tilde{u}_{jL}\bar{d}_{kR}l_{iL} - \tilde{d}_{kR}^*\bar{l}_{iR}^cu_{jL}] + \text{H.c}. 
\end{split}
\end{equation}   

The soft terms corresponding to the $R$-parity violating superpotential involving these two trilinear couplings are given by~\cite{Barbier:2004ez}
\begin{equation}
\label{eq:soft_lag}
V_{\cancel{L}}^{soft} = \frac{1}{2}A_{ijk} \tilde{L}_i\tilde{L}_j\tilde{l}_k^c + A^{\prime}_{ijk} \tilde{L}_i\tilde{Q}_j\tilde{d}_k^c + \text{H.c}.
\end{equation}
The soft coupling terms also have 9 and 27 independent parameters corresponding to $A_{ijk}$ and $A_{ijk}^{\prime}$ terms, respectively. The trilinear RPV couplings ($\lambda$ and $\lambda^{\prime}$) and the corresponding soft coupling terms ($A_{ijk}$ and $A_{ijk}^{\prime}$) are related with the physical Yukawa terms as follows
\begin{equation}
\label{eq:coupling_relation}
\begin{split}
\lambda_{ijk}& = \sum_{\alpha , \beta} e_{\alpha i} e_{\beta j} \lambda_{\alpha \beta k}^e ~~~~~~~~~~~~~~~~~~~~~~~~~~~~~~~~
\lambda_{ijk}^{\prime} = \sum_{\alpha} e_{\alpha i} \lambda_{\alpha j k}^d \\
A_{ijk} &= \sum_{\alpha , \beta} e_{\alpha i} e_{\beta j} A_{\alpha \beta k}^e ~~~~~~~~~~~~~~~~~~~~~~~~~~~~~~~~
A_{ijk}^{\prime} = \sum_{\alpha} e_{\alpha i} A_{\alpha j k}^{\prime d} 
\end{split}
\end{equation}
where $\lambda_{\alpha \beta k}^e$ and $\lambda_{\alpha j k}^d$ with $\alpha \equiv \beta = (0,1,2,3)$\footnote{Here $\lambda_{0jk}^e = \lambda_{jk}^0$ and $\lambda_{0jk}^d = \lambda_{jk}^d$.} are physical Yukawa terms and $\vec{e}_i \equiv \bigl\{e_{\alpha i}\bigl\}_{\alpha = 0,..,3}$ obeys $\vec{v} \cdot \vec{e}_i = 0$ and $\vec{e}_i \cdot \vec{e}_j = \delta_{ij}$~\cite{Barbier:2004ez}. Similarly, $A_{\alpha \beta k}^e$ and $A_{\alpha j k}^{\prime d}$ are physical terms corresponding to soft couplings. The one loop contribution to the 
\begin{figure}[!htb]
\includegraphics[scale=1]{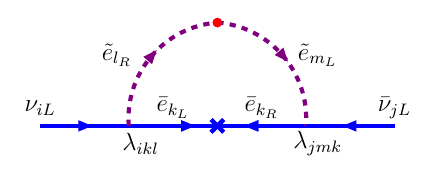} \hspace{5mm}
\includegraphics[scale=1]{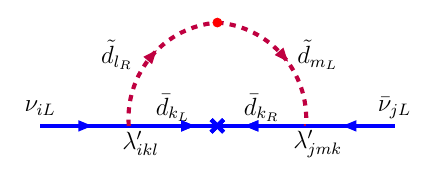}
\caption{The loop diagrams contributing to the light neutrino masses aided by $\lambda$ couplings (left) and $\lambda^{\prime}$ couplings (right) are shown here. The Majorana mass and the scalar mass insertions are represented by the cross and the dot on the respective lines for neutrino and scalar sparticles (slepton and squark), respectively.}
\label{fig:loop}
\end{figure}
light neutrino mass matrix arising from the $\lambda$, $\lambda^{\prime}$ couplings are presented in Fig.~\ref{fig:loop}. These contributions can be expressed as~\cite{Barbier:2004ez, HALL1984419, PhysRevLett.64.1705, Grossman:1998py}
\begin{equation}
\label{eq:mass1}
\begin{split}
M_{ij}^{\nu}|_{\lambda} &= \frac{1}{16\pi^2} \sum_{k,l,m} \lambda_{ikl}\lambda_{jmk} ~m_{e_{k}} ~\frac{\bigl( \tilde{m}_{LR}^{e^2} \bigl)_{ml}}{m^2_{\tilde{e}_{Rl}} - m^2_{\tilde{e}_{Lm}}}~{\rm ln}\biggl( \frac{m^2_{\tilde{e}_{Rl}}}{m^2{\tilde{e}_{Lm}}} \biggl )  + ~(i \leftrightarrow j) \\
M_{ij}^{\nu}|_{\lambda^{\prime}} &= \frac{3}{16\pi^2} \sum_{k,l,m} \lambda_{ikl}^{\prime} \lambda_{jmk}^{\prime}~ m_{d_{k}} ~\frac{\bigl( \tilde{m}_{LR}^{d^2} \bigl)_{ml}}{m^2_{\tilde{d}_{Rl}} - m^2_{\tilde{d}_{Lm}}}~{\rm ln}\biggl( \frac{m^2_{\tilde{d}_{Rl}}}{m^2{\tilde{d}_{Lm}}} \biggl ) + ~(i \leftrightarrow j)
\end{split}
\end{equation} 
The functional forms are identical except for a factor of 3 in the mass contribution coming from $\lambda^{\prime}$, which arises due to different color combinations. The slepton mixing term can be expressed in terms of coupling parameters as $\tilde{m}_{LR}^{e^2} = \bigl ( A^e_{ij} - \mu \tan\beta \lambda_{ij}^e \bigl )\frac{v_d}{\sqrt{2}}$, where all the charged lepton masses and the associated doublet and singlet mass matrices are diagonal. Similarly, for the squarks, assuming the down quark masses and their corresponding scalar doublets and singlets are diagonal, the mixing term can be expressed as $\tilde{m}_{LR}^{d^2} = \bigl ( A^d_{ij} - \mu \tan\beta \lambda_{ij}^e \bigl )\frac{v_d}{\sqrt{2}}$.
Furthermore, under certain assumptions such as degenerate sfermion masses and $A$-terms proportional to Yukawa couplings ($A_{ij}^e = A^e \lambda_{ij}^e$ and $A_{ij}^d = A^d \lambda_{ij}^d$), the equations in Eq. \ref{eq:mass1} can be reformulated as follows~\cite{Barbier:2004ez, Grossman:1998py, Joshipura:1999hr, Davidson:2000ne} 
\begin{equation}
\label{eq:mass2}
\begin{split}
M_{ij}^{\nu}|_{\lambda} \simeq \frac{1}{8\pi^2}~\frac{A^e - \mu \tan\beta}{\bar{m}^2_{\tilde{e}}} ~\sum_{k,l} \lambda_{ikl}\lambda_{jkl} ~m_{e_{k}} m_{e_{l}} \\
M_{ij}^{\nu}|_{\lambda^{\prime}} \simeq \frac{3}{8\pi^2}~\frac{A^d - \mu \tan\beta}{\bar{m}^2_{\tilde{d}}} ~\sum_{k,l} \lambda_{ikl}^{\prime}\lambda_{jkl}^{\prime} ~m_{d_{k}} m_{d_{l}}
\end{split}
\end{equation}
Here, $\bar{m}_{\tilde{e}}$ and $\bar{m}_{\tilde{d}}$ denote the scalar mass parameters corresponding to the charged lepton and down quark mass eigenstates, respectively. Although the equations in Eq.~\ref{eq:mass2} do not explicitly depend on $A$ or $A^{\prime}$, but they involve a large number of independent parameters (40), including $\lambda_{ijk}$ (9), $\lambda_{ijk}^{\prime}$ (27), $\mu$, $\tan\beta$, $\bar{m}^2_{\tilde{e}}$, and $\bar{m}^2_{\tilde{d}}$.
To simplify, one can assume that there is no strong hierarchy among $\lambda_{ijk}$ and $\lambda_{ijk}^{\prime}$ couplings for a given index $i$~\cite{PhysRevD.52.5319, Binetruy:1997sm, Borzumati:1996hd, Grossman:1998py}. Additionally, assuming models where fermion mass hierarchy is explained by flavor symmetries, contributions with $k, l = 2, 3$ dominate~\cite{Grossman:1998py} in the equations in Eq. \ref{eq:mass2}. Furthermore, setting all sfermion mass parameters ($\bar{m}_{\tilde{d}}^2$ and $\bar{m}_{\tilde{e}}^2$) as $\tilde{m}$, contributions to the light neutrino mass matrix can be expressed as~\cite{Grossman:1998py,Barbier:2004ez,Joshipura:1999hr, Davidson:2000ne, Dreiner:2022zsc}
\begin{equation}
\label{eq:mass3}
\begin{split}
M_{ij}^{\nu}|_{\lambda} \simeq \frac{1}{8\pi^2} \biggl\{ \lambda_{i33} \lambda_{j33} ~\frac{m_{\tau}^2}{\tilde{m}} + \Bigl(\lambda_{i23}\lambda_{j32} +\lambda_{i32}\lambda_{j23}\Bigl)~ \frac{m_{\mu}m_{\tau}}{\tilde{m}} + \lambda_{i22}\lambda_{j22} \frac{m_{\mu}^2}{\tilde{m}} \biggl\} \\
M_{ij}^{\nu}|_{\lambda^{\prime}} \simeq \frac{3}{8\pi^2} \biggl\{ \lambda_{i33}^{\prime} \lambda_{j33}^{\prime} ~\frac{m_{b}^2}{\tilde{m}} + \Bigl(\lambda_{i23}^{\prime}\lambda_{j32}^{\prime} +\lambda_{i32}^{\prime}\lambda_{j23}^{\prime}\Bigl)~ \frac{m_{s}m_{b}}{\tilde{m}} + \lambda_{i22}^{\prime}\lambda_{j22}^{\prime} \frac{m_{s}^2}{\tilde{m}} \biggl\}
\end{split}
\end{equation} 

In this pair of mass equations, a natural hierarchy emerges in fermion masses, such as $\frac{m_{\tau}^2}{m_{\mu}^2} \sim 300$ and $\frac{m_{b}^2}{m_s^2} \sim 600$~\citep{Barbier:2004ez,Davidson:2000ne,Dreiner:2022zsc}. Consequently, our focus lies on the contributions stemming from fermions with significant masses, namely, the $b$ quark and $\tau$ lepton. In this context, various models with diagonal and off-diagonal RPV trilinear couplings have been explored in existing literature~\cite{Barbier:2004ez, Dreiner:2022zsc}. In our study, we consider a simple diagonal model presented in~\cite{Dreiner:2022zsc}, and by combining these two diagonal contributions, the final light neutrino mass matrix can be expressed as
\begin{equation}
\label{eq:mass4}
M_{\nu}|_{\lambda+\lambda^{\prime}} = \frac{1}{8\pi^2\tilde{m}} \Bigr[\lambda_{i33} \lambda_{j33}~m_{\tau}^2 + 3\lambda^{\prime}_{i33} \lambda_{j33}^{\prime}~m_b^2 \Bigr]
= \frac{1}{8\pi^2\tilde{m}} M_{ij}^{\nu}
\end{equation}
where the matrix $M_{ij}^{\nu}$ has the form  

\begin{equation}
\label{eq:mass5}
M_{ij}^{\nu} = 
\setlength\arraycolsep{8pt}
\begin{pmatrix}
\lambda_{133}^2m_{\tau}^2 + 3\lambda_{133}^{\prime^2}m_b^2 & \lambda_{133}\lambda_{233}m_{\tau}^2+3\lambda_{133}^{\prime}\lambda_{233}^{\prime}m_b^2 & 3\lambda_{133}^{\prime}\lambda_{333}^{\prime}m_b^2  \\[8pt]
\lambda_{233}\lambda_{133}m_{\tau}^2+3\lambda_{233}^{\prime}\lambda_{133}^{\prime}m_b^2 & \lambda_{233}^2m_{\tau}^2 + 3\lambda_{233}^{\prime^2}m_b^2 & 3\lambda_{233}^{\prime}\lambda_{333}^{\prime}m_b^2 \\[8pt]
3\lambda_{333}^{\prime}\lambda_{133}^{\prime}m_b^2 & 3\lambda_{333}^{\prime}\lambda_{233}^{\prime}m_b^2 & 3\lambda_{333}^{\prime^2}m_b^2
\end{pmatrix}
\end{equation}
Due to the asymmetry of the $\lambda_{ijk}$ coupling, the absence of a $\lambda_{333}$ term is evident in this mass matrix. 
The third neutrino gets mass at the leading order, and it can be approximated as~\cite{Barbier:2004ez} 
\begin{equation}
\label{eq:third_neutrino_mass}
m_{\nu_{3}} = \frac{3m_{b}^2}{8\pi^2\tilde{m}} \sum_{i} {\lambda^{\prime}}^2_{i33}
\end{equation}  

\section{Constraints from neutrino oscillation, flavor physics and Higgs data}
\label{sec:observable}
We have considered two mass square differences, namely, $\Delta m_{21}^2$ and $|\Delta m_{31}^2|$ ($\Delta m_{i1}^2 = m_i^2 - m_1^2$, where $i=2,3$) along with three mixing angles ($\theta_{13}$, $\theta_{12}$ and $\theta_{23}$) obtained from global fit of neutrino oscillation data. The CP violating phase $\delta_{CP}$ has not been considered in our analysis. There are still large error bars associated with measurements of CP-violating phases in neutrino oscillation experiments. For example, in the normal hierarchical scenario, the phase is consistent with $\pi$ within $1\sigma$ of the best-fit value \cite{deSalas:2020pgw}, which makes the scenario consistent with zero CP phase. The data corresponding to inverted hierarchy also comes with a significantly large uncertainty \cite{deSalas:2020pgw}. Adding these observables does not restrict our parameter regions any further. On the other hand, to address nonzero values of the phase angle, one is required to add additional input parameters, which makes the numerical computation even more time consuming. Hence, we have calculated the mixing angles from the PMNS matrix by putting $\delta_{CP} = 0$~\cite{Donini:1999jc, Akhmedov:1999uz, Giganti:2017fhf}. Several groups have performed the global analysis of the neutrino oscillation results coming from the different experiments \cite{Esteban:2016qun,deSalas:2020pgw, Esteban:2020cvm}. We have considered the values of neutrino oscillation parameters as listed in Table.~\ref{tab:neutrino_obs}, which are reported in the Ref.~\cite{deSalas:2020pgw}. We have also ensured that the sum total of light neutrino masses satisfies the limit arising from cosmological data~\cite{deSalas:2020pgw} as mentioned in the last row of Table.~\ref{tab:neutrino_obs}. \\
\begin{table}[!htb]
\hrule\hrule 
\vspace*{3mm} \hspace*{8mm} \textbf{Neutrino oscillation} \hspace{4.9cm} \textbf{Best-fit value $\pm~1\sigma$} \\
\hspace*{20mm} \textbf{parameter} \hspace{5.9cm} \textbf{NH} \hspace{1.6cm} \textbf{IH} \\  
\hrule\hrule \vspace{3mm}
\hspace*{16mm} $\Delta m^2_{21}$[$10^{-5}$eV$^2$] \hspace{5.4cm} 7.50$^{+0.22}_{-0.20}$  \hspace{0.57cm} 7.50$^{+0.22}_{-0.20}$ \vspace{3mm}\\
\hspace*{16mm}  $|\Delta m^2_{31}|$[$10^{-3}$eV$^2$]  \hspace{5.2cm} 2.55$^{+0.02}_{-0.03}$ \hspace{0.5cm} 2.45$^{+0.02}_{-0.03}$ \vspace{3mm}\\
\hspace*{28mm} $\theta_{12}/^{\circ}$  \hspace{6.1cm} 34.3$^{+1.0}_{-1.0}$ \hspace{0.65cm} 34.3$^{+1.0}_{-1.0}$ \vspace{3mm}\\
\hspace*{28mm}  $\theta_{13}/^{\circ}$ \hspace{6.1cm} 8.53$^{+0.13}_{-0.12}$ \hspace{0.5cm} 8.58$^{+0.12}_{-0.14}$ \vspace{3mm}\\
\hspace*{28mm}  $\theta_{23}/^{\circ}$ \hspace{6.0cm} 49.26$^{+0.79}_{-0.79}$ \hspace{0.3cm} 49.46$^{+0.60}_{-0.97}$ \vspace{3mm}\\
\hspace*{22mm}  $\sum m_{\nu_i}$[eV] \hspace{6.0cm} $< 0.12$ \hspace{0.8cm} $<0.15$ \\
\hrule 
\hrule \vspace{3mm}
\caption{The best-fit value along with $1\sigma$ value of neutrino oscillation parameters taken from the global fit~\cite{deSalas:2020pgw} of different results coming from various experiments are listed in this table. Here NH and IH refer to the normal hierarchy and inverted hierarchy scenarios respectively.}
\label{tab:neutrino_obs}
\end{table}

Constraints on the LLE and LQD couplings also arise from nonobservation of neutrinoless double beta decay \cite{Babu:1995vh,Hirsch:1995vr,Hirsch:1995ek,Allanach:2014lca,Allanach:2014nna}. The existing limits on the lifetime is $\tau > 1.07\times 10^{26}$ years as quoted by the KamLAND-Zen experiment \cite{Jones:2021cga,KamLAND-Zen:2016pfg}, which restricts the $\lambda$ and $\lambda^{\prime}$ couplings to be too large \cite{Barbier:2004ez}\footnote{From our analysis we find that the neutrino oscillation data is more constraining.}. 

Some SUSY parameters (e.g., $\mu$, $\tan\beta$) affect the neutrino and Higgs sectors. Therefore, in addition to neutrino data, we have also considered updated data corresponding to the Higgs sector in the form of the SM Higgs mass ($m_h$) and its coupling strengths. A statistical combination of the updated measurements of the Higgs mass by the ATLAS and CMS Collaborations puts the mass at 125.09~$\pm$~0.21 (stat.)~$\pm$ 0.11 (syst.)GeV~\cite{CMS:2012qbp,ATLAS:2012yve,ATLAS:2015yey}. Considering the theoretical uncertainty within the SUSY framework for the Higgs mass calculation, we use a $\pm$3 GeV~\cite{Allanach:2004rh} window around the combined best-fit value of $m_{h}$. 
The Higgs boson coupling strength modifiers for each decay channel $i$ are defined as $\kappa_i^2 = \frac{\Gamma_i}{\Gamma^{SM}_i}$, where $\Gamma_i$ and $\Gamma^{SM}_i$ are the BSM and SM contribution to the decay width respectively. Both the ATLAS and CMS Collaborations measure these coupling strength modifiers and present the results with their corresponding uncertainties. The most updated results thus far are quoted in Ref.~\cite{ATLAS:2021vrm, CMS-PAS-HIG-19-005}. In this analysis, we only consider the results obtained by the CMS Collaboration\cite{CMS-PAS-HIG-19-005} and summarize them in Table.~\ref{tab:Higgs_coupling}\footnote{Note that the ATLAS Collaboration results\cite{ATLAS:2021vrm} are very similar and will lead to similar posterior distributions of the input parameters.}. The effective Hiigs-gluon-gluon coupling strength is quite sensitive to the SUSY-breaking scale ($M_{\text{SUSY}}$). However, varying the SUSY breaking scale would require us to vary the trilinear top Yukawa coupling ($A_t$) and the stop masses as well in addition to $\tan\beta$ and $\mu$ in order to ensure the 125 GeV Higgs mass. Therefore, to avoid this complexity and to reduce the number of free parameters for fast computation, we avoid this scenario as our main focus is on the neutrino sector.
\begin{table}[!htb]
\hrule\hrule 
\vspace*{3mm} \hspace*{8mm} \textbf{Coupling strength modifiers} \hspace{3.0cm} \textbf{Best-fit value $\pm~1\sigma$} \\ [0.1ex] 
\hrule\hrule \vspace{3mm}
\hspace*{32mm} $\kappa_z$ \hspace{7cm} $0.96^{+0.07}_{-0.07}$ \vspace{2mm}\\
\hspace*{32mm} $\kappa_w$ \hspace{7cm} $1.11^{+0.14}_{-0.09}$ \vspace{2mm}\\
\hspace*{32mm} $\kappa_b$ \hspace{7cm} $1.18^{+0.19}_{-0.27}$ \vspace{2mm}\\
\hspace*{32mm} $\kappa_{\tau}$ \hspace{7cm} $0.94^{+0.12}_{-0.12}$ \vspace{2mm}\\
\hspace*{32mm} $\kappa_{\mu}$ \hspace{7cm} $0.92^{ +0.55}_{ -0.87}$ \vspace{2mm}\\
\hspace*{32mm} $\kappa_t$ \hspace{7cm} $1.01^{+0.11}_{-0.11}$ \vspace{2mm}\\
\hspace*{32mm} $\kappa_{\gamma}$ \hspace{7cm} $1.01^{+0.09}_{-0.14}$ \vspace{1mm}\\
\hrule 
\hrule \vspace{3mm}
\caption{The best-fit along with $1\sigma$ values of Higgs boson coupling strength modifiers obtained by the CMS Collaboration at $\sqrt{s} = 13$ TeV energy with luminosity $\mathcal{L}= 137$ fb$^{-1}$~\cite{CMS-PAS-HIG-19-005}.}
\label{tab:Higgs_coupling}
\end{table} 

There are different RPC and RPV SUSY contributions to the flavor violating decays like $B \rightarrow X_s \gamma$, $B_s \rightarrow \mu^+ \mu^-$ and $B \rightarrow \tau \nu$. So, the limit on the branching ratios of these decays can constrain the parameter space of both RPC and RPV SUSY. There are several RPC contributions to the branching ratio of $B \rightarrow X_s \gamma$ decay like gaugino-down type squark, chargino-up type squark, charged Higgs-top quarks, etc.~\cite{Bertolini:1994cv, deCarlos:1996yh}. In the RPV SUSY the main contribution is coming from $\lambda^{\prime}$ coupling and the contributing coupling products are $\lambda_{i2j}^{\prime} \lambda_{i3j}^{\prime}$ and $\lambda_{ij3}^{\prime} \lambda_{ij2}^{\prime}$~\cite{deCarlos:1996yh, Barbier:2004ez, Dreiner:2013jta} corresponding to the $B \rightarrow X_s \gamma$ decay branching ratio. Similarly, for $B_s \rightarrow \mu^+ \mu^-$ decay, the RPC SUSY contributions are $Z^0$ penguin and box diagrams mediated through charged Higgs boson~\cite{Babu:1999hn}. The RPV contribution to this branching ratio depends on both the couplings $\lambda$ and $\lambda^{\prime}$ and the related coupling products are $\lambda_{122}\lambda_{i23}^{\prime}$, $\lambda_{i22}\lambda_{i32}^{\prime}$ and $\lambda_{2j2}^{\prime}\lambda_{2j3}^{\prime}$~\cite{Barbier:2004ez, Dreiner:2013jta}. There will be a radiative correction to the Higgs boson mass and couplings due to the contributions from charginos, neutralinos, and sfermions, and this change will contribute to the branching ratio of $B \rightarrow \tau \nu$ decay. In the RPV SUSY, there will be additional four-fermion interactions due to the exchange of sleptons and squarks, and this will affect the leptonic decays of $B$ meson~\cite{Baek:1999ch,Barbier:2004ez}. The relevant coupling product for this decay is $\lambda_{313}^{\prime}\lambda_{233}$~\cite{Barbier:2004ez}. As we have considered only $\lambda_{i33}$ and $\lambda_{i33}^{\prime}$ for our analysis, then for all these flavor violating decay, only RPC contribution will be effective. We have taken the value of $B_s \rightarrow \mu^+ \mu^-$ as (3.09$^{+0.46 + 0.15}_{-0.43 - 0.11}$)$ \times 10^{-9}$ which is provided by the LHCb Collaboration~\cite{LHCb:2021vsc}. The world average of corresponding to $B \rightarrow X_s  \gamma$ is given as (3.32 $\pm$ 0.15)$\times 10^{-4}$ in the literature~\cite{HFLAV:2019otj}. Similarly, we have also taken into account the branching ratio of $B \rightarrow \tau \nu$ as (1.06 $\pm$ 0.19)$\times 10^{-4}$~\cite{HFLAV:2019otj}.


\section{Constraints on sparticles from direct searches at LHC}
\label{sec:constraints_mass} 
While fitting the neutrino data, we need to satisfy the current exclusion limit on the sparticle masses coming from the searches using the LHC data. Both the ATLAS and CMS Collaborations have studied a wide range of final states in search of the sparticles. The results coming from these direct searches using Run-I and Run-II data are summarized in Ref.~\cite{atlas_web, cms_web}. These experimental bounds on sparticle masses are derived with certain simplified assumptions like the presence of only one nonzero RPV coupling, 100\% branching ratio to a particular decay mode, etc. In this analysis, as we have considered a model with nonzero values of both the lepton number violating trilinear couplings ($\lambda$ and $\lambda^{\prime}$), we focus on the corresponding exclusion limits on sparticle masses. In a recent review \cite{Choudhury:2024ggy}, the authors have discussed in detail about the current exclusion limits on different sparticles with respect to several final state searches for RPV couplings (see Sec.6 of Ref.\cite{Choudhury:2024ggy}). Notably, when the RPV couplings are very small, sparticles are expected to follow RPC SUSY decay modes, and we also include the exclusion limits associated with these RPC decays.

\subsection{Exclusion limits on strong sector}
\label{sec:strong}

\textbf{Gluino search:}
In the RPC framework, the ATLAS and CMS Collaborations have set the 
most stringent bounds on the masses of gluinos up to $\sim$ 
2.0-2.3 TeV~\cite{CMS:2019zmd, ATLAS:2020syg, ATLAS:2021twp, CMS:2021beq, CMS:2019ybf, CMS:2020cpy, CMS:2022idi, CMS:2020bfa} for different decay modes, with neutralino masses up to 600 GeV. Also, 
 several searches focused on the LQD or LLE coupling using Run-I and Run-II data have been conducted by both the ATLAS~\cite{ATLAS:2021fbt, ATLAS:2023afl,ATLAS:2018rns, ATLAS:2021yyr} and CMS~\cite{CMS:2020cpy,CMS:2016zgb} groups where gluino decays indirectly. 
For $\lambda^{\prime}$ coupling the considered decay mode is   
$\gl \to q \bar q\lspone \to q \bar q (q\bar{q}l/\nu) $, where $q$, $l$, and $\nu$ refer to light quarks, leptons, and neutrinos, respectively. ATLAS and CMS have excluded gluino masses up to $\sim$ 2.1 - 2.2 TeV for bino-type LSP with $m_{\lspone} < 1000$ GeV using LHC Run-II data~\cite{ATLAS:2021fbt,CMS:2020cpy,
ATLAS:2023afl}. Using the full Run-II dataset, 
the ATLAS Collaboration has updated the lower bounds on gluino masses up to 2.5 TeV (2.0 TeV) for $\lambda_{121/122}$  ($\lambda_{133/233}$ ) type LLE scenarios ~\cite{ATLAS:2021yyr}.

\noindent
\textbf{Squark search:} 
Similar to the gluino searches, both the ATLAS and CMS have conducted squark searches using several final states within the framework of RPC and RPV SUSY models. In RPC MSSM scenarios, these different analyses constrain the lightest stop quark ($\st1$) mass up to 1.0 - 1.3 TeV ~\cite{CMS:2019ybf, CMS:2019ysk, CMS:2021beq, ATLAS:2020dsf, ATLAS:2020xzu, ATLAS:2021hza}, the lightest sbottom quark ($\sbot1$) mass upto 1.15 - 1.28 TeV~\cite{CMS:2019zmd, CMS:2019ybf, ATLAS:2021yij}, first two generation light squarks mass upto 1.25-1.85 TeV~\cite{ATLAS:2018nud, CMS:2019zmd, CMS:2019ybf, CMS:2021eha, ATLAS:2020syg, ATLAS:2021twp, ATLAS:2022zwa, ATLAS:2021kxv, ATLAS:2023afl} for a massless neutralino. In the RPV scenarios, the exclusion limit on the $\st1$  mass, from stop NLPS pair production,  has reached up to $\sim$ 820 to 1020 GeV depending on LLE/LQD couplings ~\cite{CMS:2013pkf}. For the production of degenerate first two generations light squark (NLSP), the limits are set at 1.6 TeV ~\cite{CMS:2016zgb}. For direct decay of the stop squark, the ATLAS experiment has excluded $\st1$ masses up to 1.9 TeV, 1.8 TeV, and 800 GeV for stop decays to $be$, $b\mu$, and $b\tau$ respectively, with a 100\% branching ratio~\cite{ATLAS:2017jvy, ATLAS:2024zkx}. 

\subsection{Exclusion limits on electroweak sector}
\label{sec:electroweak}
The limits on the electroweak sector are relatively weaker. For winolike lighter chargino ($\chonepm$) or second lightest neutralino ($\lsptwo$) and binolike LSP ($\lspone$), the most stringent limits come from $\chonepm \lsptwo$ pair production. Depending on the decay modes of NLSP (e.g., slepton mediated, WZ or Wh mediated), the ATLAS and CMS groups have excluded  $\mchonepm$ (=$\mlsptwo$) in the range 1 - 1.2  TeV~\cite{ATLAS:2019lff, ATLAS:2018ojr, ATLAS:2022nrb, ATLAS:2020pgy, ATLAS:2021yqv, ATLAS:2022zwa, ATLAS:2021moa, CMS:2022sfi, CMS:2021cox, CMS:2024gyw, atlas_web, cms_web}\footnote{For Higgsino pair production this limits reduce to around 200 GeV \cite{ATLAS:2024qxh, ATLAS:2024tqe} in RPC MSSM.}. The CMS Collaboration has also looked for multilepton final state in the context of different RPV leptonic (LLE) and semileptonic (LQD) couplings and excluded wino and Higgsino like LSP in the 300-900 GeV mass range  \cite{CMS:2016zgb}. Considering winolike  $\chonepm \lsptwo$ pair production and decay of LSP via $\lambda_{121/122}$  ($\lambda_{133/233}$ ) type LLE couplings the ATLAS group has excluded $\chonepm$ mass upto 1.6 TeV (1.13) TeV~\cite{ATLAS:2021yyr}. 
It may be noted that for a lighter $\lspone$, the limits get reduced drastically, and for our chosen $\mlspone$ (300 GeV), the upper limits on $\mchonepm$ becomes $\sim$ 1025 GeV for models with nonzero $\lambda_{133/233}$ couplings.
 With this same analysis, ATLAS has imposed lower bounds on slepton masses up to 1.2 TeV (860 GeV) for  $\lambda_{121/122}$  ($\lambda_{133/233}$) couplings. It may be noted that in RPC SUSY models, where a slepton decays to a lepton and an LSP, the first two generations of sleptons ($\tilde{e}$ and $\tilde{\mu}$) are excluded up to 700 GeV for a massless neutralino ($\lspone$) ~\cite{ATLAS:2019lff,CMS:2020bfa}, while the limit for the stau is much weaker. 

The MSSM Higgs sector,  at the tree level, 
 is essentially described by two parameters  - 
pseudoscalar mass ($M_A$) and tan$\beta$. The most stringent bound in the 
$M_A$~-~tan$\beta$ plane comes from the heavy Higgs searches 
where the neutral heavy Higgs decays via  $H/A \rightarrow \tau^+ \tau^-$ \cite{CMS:2018rmh, ATLAS:2020zms}. The LHC Run-II data has excluded 
$M_A$ up to $1.0$ ($1.5$) TeV for $\tan\beta < 8$ ($21$).

\section{Details of MCMC analysis setup}
\label{sec:analysis_details}
In this section, we discuss the details of the analysis setup. The analytical calculations for the corresponding trilinear RPV model starting from the Lagrangian are performed through \texttt{SARAH}~\cite{Staub:2008uz,Staub:2010jh,Staub:2015kfa} which provides all the files necessary for the required numerical calculations which are subsequently performed by \texttt{SPheno}~\cite{Porod:2003um,Porod:2011nf}. All the flavor observables mentioned in Sec.~\ref{sec:observable} are obtained from \texttt{FlavorKit}~\cite{Porod:2014xia}, which is implemented within the framework of \texttt{SPheno}. \texttt{SPheno} employs two-loop and one loop calculations for Higgs mass and all other particle masses, respectively. \texttt{FlavorKit} calculates the flavor physics observables up to one loop level. Our primary objective in this analysis is to find out the allowed parameter space in the form of marginal posterior distributions that can fit all the observables while satisfying all constraints detailed in Sections \ref{sec:observable}  and \ref{sec:constraints_mass}. To achieve this, we employ the MCMC method using the publicly available \texttt{emcee} code \cite{foreman2013emcee}. Our approach involves conducting a likelihood analysis by minimizing the chi-square function defined as
\begin{equation}
\mathcal{L} = \frac{\chi^2}{2} = \frac{1}{2}\sum_{i=1}^{n_{\rm obs}} \left[ \frac{X_i^{\rm obs} - X_i^{\rm th}}{\sigma_i} \right]^2,
\end{equation}
where $\mathcal{L}$ represents the negative of log-likelihood, related to the likelihood by $L \propto \exp(-\mathcal{L})$. Here, $X_i^{\rm obs}$ denotes the experimentally observed values of $X_i$, each with corresponding uncertainties $\sigma_i$, and $X_i^{\rm th}$ represents the theoretically calculated value of each observable using our RPV model. In this analysis, we have a total of 16 observables.
The number of free parameters is different for different models, which we will discuss in the respective sections, and according to that the degrees of freedom (\texttt{dof}) will also change. 
We adopt flat priors for each parameter but draw samples using a Gaussian distribution function. To ensure the comprehensive exploration of the parameter space and to gather sufficient data for drawing $1\sigma$, $2\sigma$ and $3\sigma$ contours, we employ 500 walkers ($n_{\rm walker}$) and 400 steps ($n_{\rm step}$) corresponding to each walker. Depending upon the number of cores ($n_{core} \sim$ 16-20) we use for the analysis, the number of total generated samples ($n_{\rm walker} \times n_{\rm step} \times n_{\rm core} = 200000\times n_{core}$) will be changed. We have also conducted an auto-correlation analysis to ensure that each chain meets the convergence criterion satisfactorily.

In the next section we will focus on estimating the allowed parameter space of different RPV scenarios by calculating the marginalized posterior distribution of various input parameters. 
The MCMC analysis also searches for the minimum $\chi^2$ ($\chi^2_{\text{min}}$)
and we calculate the \texttt{p-value} for this $\chi^2$ statistics.
The \texttt{p-value} is defined as $p=P(\chi^2 \ge \chi^2_{\text{min}} | H_0$)~\cite{ParticleDataGroup:2024cfk} where $H_0$ is the null hypothesis, typically stating that the model or hypothesis fits the data well. This value can be calculated using the formula $\texttt{p}= \int_{\chi^2_{\text{min}}}^{ \infty} f(\chi^2,n_d)~d\chi^2$, where $f(\chi^2,n_d)$ and $n_d$ refer to the $\chi^2$ statistics and number of degrees of freedom respectively~\cite{ParticleDataGroup:2024cfk}. A low \texttt{p-value} (typically $< 0.05$) indicates that the model does not fit the observed data well, suggesting the need to reconsider the model or the parameters of the model. On the other hand, a high \texttt{p-value} indicates that the model is consistent with the observed data.

\section{RPV models with a bino-LSP}
\label{sec:model_bino}
For a generic trilinear RPV model, the number of input parameters is quite large, and scanning the full parameter space is computationally very expensive. 
As discussed in Sec.~\ref{sec:model}, the dominant contributions to neutrino masses come from the two $\lambda_{i33}$ ($i = 1,2$) and three $\lambda_{i33}^{\prime}$ ($i = 1,2,3$) coupling parameters~\cite{Barbier:2004ez, Grossman:1998py, Joshipura:1999hr, Davidson:2000ne}. The other relevant parameters in this context are $\mu$, $\tan\beta$, $A_d$, $A_e$, and the masses of $\tilde{l}$ and $\tilde{q}$. In this work, we sequentially analyze a few simplified models considering the bino-like lightest neutralino ($\lspone$) or top squark ($\st1$) as the LSP. We have analyzed the model with only LQD type couplings (\texttt{Bino$_{\lambda^\prime}$-model}) by 
varying five input parameters 
($\lambda_{133}^{\prime}$, $\lambda_{233}^{\prime}$, $\lambda_{333}^{\prime}$, $\mu$, and $\tan\beta$) and the model with both LLE and LQD couplings  (\texttt{Bino$_{\lambda \lambda^\prime}$-model}) by varying seven input 
($\lambda_{133}$, $\lambda_{233}$, $\lambda_{133}^{\prime}$, $\lambda_{233}^{\prime}$, $\lambda_{333}^{\prime}$, $\mu$, and $\tan\beta$) parameters. 
The soft coupling parameters ($A_{i33}$ or $A_{i33}^{\prime}$) are  chosen as zero in these three scenarios. Since the soft couplings corresponding to the nonzero $\lambda$ and $\lambda^{\prime}$ couplings do not have any direct contribution to the light neutrino masses and mixings, we chose them to be zero for all these models\footnote{To check for higher order corrections, we have also considered a \texttt{Bino$_{\lambda \lambda^\prime}^{A A^\prime}$-model} with nonzero soft RPV coupling parameters and presented the results in Appendix~\ref{sec:appendix2}.}.

In this section, we probe the above-mentioned two models to fit the neutrino oscillation, Higgs, and flavor data by considering the lightest neutralino ($\lspone$) as the LSP which is bino type. To avoid the current exclusion limits from the LHC, as discussed in Sec.~\ref{sec:constraints_mass}, we fix the corresponding soft masses for bino ($M_1$) at 300 GeV, wino ($M_2$) at 1200 GeV, gluino ($M_3$) at 3 TeV, all three generation squark ($m_{\tilde{q}}$) at 3 TeV, all three generation slepton ($m_{\tilde{l}}$) at 2 TeV and $M_A$ at 3 TeV. After some preliminary scans, we have kept the third generation trilinear coupling parameter $A_t$ = -5.0 TeV for this scenario so that the SM-like Higgs mass remains around 125 GeV. The ranges of both the varying and fixed values of all input parameters used in our analysis are presented in Table.~\ref{tab:parameter_range}. We have chosen our ranges exhaustively to ensure that no allowed region is overlooked. For each model, we first obtain the  $\chi^2_{\text{min}}$, $\chi^2_{\text{min}}/\texttt{dof}$ and \texttt{p-value}. If the \texttt{p-value} and the $\chi^2_{\text{min}}/\texttt{dof}$ are satisfactory, then we proceed to discuss the allowed parameter space and present the 2D marginalized posterior distribution. 
\begin{table}[!htb]
\small
  \centering
  \begin{tabular}{||c||c|c|c|c||}
  \hline\hline
  \multirow{4}{*}{\rotatebox{90}{Parameter}} & \multicolumn{3}{c||}{Value/range of parameters in different model } \\
  \cline{2-4}
   &  \multicolumn{2}{c|}{Models with bino LSP} & Model with stop LSP \\ 
  \cline{2-4}
  & Only $\lambda_{i33}^{\prime}$ & $\lambda_{i33}$ \& $\lambda_{i33}^{\prime}$ & $\lambda_{i33}$ \& $\lambda_{i33}^{\prime}$ \\
  &  &  &  \\
  \hline

    &     \texttt{Bino$_{\lambda^\prime}$-model} & \texttt{Bino$_{\lambda \lambda^\prime}$-model}   & \texttt{Stop$_{\lambda \lambda^\prime}$-model} \\
  
  \hline\hline
  $|\lambda_{i33}|$ & - & [0-0.001] & [0-0.001]\\
  \hline 
  $|\lambda_{i33}^{\prime}|$  & [0-0.001] & [0-0.001] & [0-0.001]\\
  \hline
  $\mu$  & [800-3000] & [800-3000] &  $m_{\tilde{q}_{3L}}+1000$ \\
  \hline
  $\tan\beta$ & [1-60] & [1-60] & [1-60] \\
  \hline 
  $m_{\tilde{q}_{3L}}$  & 3000 & 3000 & [1000-3000] \\
  \hline
  $M_1$  & 300 & 300 & $m_{\tilde{q}_{3L}}+1000$ \\
  \hline
  $M_2$  & 1200 & 1200 & $m_{\tilde{q}_{3L}}+1000$ \\
  \hline
  $M_3$  & 3000 & 3000 & 3000 \\
  \hline
  $M_A$ & 3000 & 3000 & 3000 \\
  \hline
  $A_t$  & -5000 & -5000 & 4500 \\
  \hline
  $m_{\tilde{l}}$  & 2000 & 2000 & $m_{\tilde{q}_{3L}}+500$ \\
  \hline
  $m_{\tilde{q}}$ & 3000 & 3000 & 3000 \\ 
  \hline
  \end{tabular}
  \caption{Ranges and fixed values of all the input parameters considered in our analysis corresponding to different models. All the parameters are in GeV unit except $\tan\beta$, $\lambda$ and $\lambda^{\prime}$. Here $i=1,2$ ($i=1,2,3$) for $\lambda_{i33}$ ($\lambda_{i33}^{\prime}$). Also $m_{\tilde{l}}$ refers to all the three generation sleptons (both L-type and R-type) and $m_{\tilde{q}}$ represents all the generations of squark (both L-type and R-type) except the third generation L-type squarks which are represented by $m_{\tilde{q}_{3L}}$. }
	\label{tab:parameter_range}
\end{table}

\vspace{5mm}
\noindent
\textbf{Model with $\lambda^{\prime}$ coupling only  (\texttt{Bino$_{\lambda^\prime}$-model}): }Since it is quite obvious from Equation~\ref{eq:mass4} that only nonzero $\lambda_{i33}$ ($i$=1,2) cannot be used to fit neutrino oscillation data, first we considered the model with nonzero $\lambda_{i33}^{\prime}$ ($i=1,2,3$) couplings keeping the corresponding soft coupling parameters and $\lambda_{i33}$  couplings zero. The leading order contribution to light neutrino masses arises from quark-squark loops. From Equation~\ref{eq:mass4}, one can see that if only $LQD$ type RPV trilinear couplings are nonzero, the mass matrix at the leading order can be approximated as $M_{\nu}|_{\lambda^{\prime}} = \frac{3}{8\pi^2\tilde{m}}~\lambda_{i33}^{\prime} \lambda_{j33}^{\prime}~m_{b}^2 $.  
The degree of freedom for this scenario is 11, as we have 5 free parameters here and 16 observables. We observe that the neutrino oscillation data cannot be fitted with this simplified model. It can explain all the neutrino oscillation observables other than $\Delta m_{21}^2$. In this analysis, we have assumed the light neutrino masses align in the normal hierarchy. Consequently, the $\chi^2_{\text{min}}/\texttt{dof}\sim 100$ is on the larger side. Hence, we are not going to discuss this scenario any further. 
\subsection{Model with both $\lambda$ and $\lambda^{\prime}$ couplings (\texttt{Bino$_{ \lambda \lambda^\prime}$-model})}
\label{sec:both_coupling}
We have already discussed that only any one type of coupling cannot explain all the neutrino observables. We need to consider both couplings together~\cite{Barbier:2004ez, PhysRevD.52.5319, Borzumati:1996hd, Dreiner:2022zsc, Drees:1997id}. The parameter space explored here is mentioned in the Table.~\ref{tab:parameter_range}. The ranges of 7 input parameters along with other fixed choices for \texttt{Bino$_{ \lambda \lambda^\prime}$-model}  are tabulated in the fourth column of Table.~\ref{tab:parameter_range}. Now, we proceed to discuss results corresponding to this model obtained for both the neutrino mass hierarchy.
\subsubsection{Normal Hierarchy scenario} 
\label{sec:nh}
In the normal hierarchy, the third neutrino mass eigenstate ($\nu_3$) is the heaviest and is predominantly $\tau$-flavored. The second neutrino mass eigenstate ($\nu_2$) has a slightly lower mass and is an almost equal mixture of all three lepton flavors. The lightest mass eigenstate ($\nu_1$) is primarily $e$-flavored.
\begin{table}[!htb]
\begin{center}
	\begin{tabular}{||c|c|c||c|c||}
	\hline\hline
	 \multicolumn{3}{||c||}{Input parameters} & \multicolumn{2}{c||}{Best-fit observable}	\\
	\hline
     Parameter & Best-fit point & Mean value$^{+1\sigma}_{-1\sigma}$   & Observable & Best-fit \\  
	\hline\hline
	\multirow{2}{*}{$\lambda_{133}\times10^{-4}$} & \multirow{2}{*}{1.71} & \multirow{2}{*}{1.81$^{+0.14}_{-0.18}$}  & $\Delta m_{21}^2$[eV$^2$] & 7.49$\times10^{-5}$  \\
	\cline{4-5}
	 &  &  & $\Delta m_{31}^2$[eV$^2$] & 2.55$\times10^{-3}$ \\
	\hline
	\multirow{2}{*}{$\lambda_{233}\times10^{-4}$} & \multirow{2}{*}{2.52} & \multirow{2}{*}{2.64$^{+0.19}_{-0.24}$}  & $\theta_{13}$ & 8.53$^{\circ}$  \\
	\cline{4-5}
	& & & $\theta_{12}$ & 34.06$^{\circ}$ \\
	\hline
	\multirow{2}{*}{$\lambda_{133}^{\prime}\times10^{-5}$} & \multirow{2}{*}{-7.61} & \multirow{2}{*}{-7.70$^{+2.90}_{-0.74}$}  & $\theta_{23}$ & 49.30$^{\circ}$ \\
	\cline{4-5}
	 &  &  & $m_h$ [GeV] & 124.88\\
	\hline
	\multirow{2}{*}{$\lambda_{233}^{\prime}\times10^{-5}$} & \multirow{2}{*}{-7.65} & \multirow{2}{*}{-8.17$^{+0.83}_{-1.99}$}  & $\mathcal{B}r(B \rightarrow X_s \gamma)$ & 3.15$\times10^{-4}$  \\
	\cline{4-5} 
	& & & $\mathcal{B}r(B_s \rightarrow \mu^+ \mu^-)$ & 3.18$\times10^{-9}$  \\
	\hline
	\multirow{2}{*}{$\lambda_{333}^{\prime}\times10^{-4}$} & \multirow{2}{*}{-1.34} & \multirow{2}{*}{-1.40$^{+0.11}_{-0.08}$}  & $\mathcal{B}r(B \rightarrow \tau \nu)$ & 1.24$\times10^{-4}$  \\
	\cline{4-5}
	 &  & & $\kappa_z$ & 1.0  \\
	\hline
	\multirow{2}{*}{$\mu$ [GeV]} & \multirow{2}{*}{1996} & \multirow{2}{*}{1820$^{+225}_{-300}$}  & $\kappa_w$ & 1.0  \\
	\cline{4-5}
	& &  & $\kappa_b$ & 1.002 \\
	\hline
	\multirow{2}{*}{$\tan\beta$} & \multirow{2}{*}{6.68} & \multirow{2}{*}{6.16$^{+2.35}_{-1.36}$}  & $\kappa_{\tau}$ & 1.002 \\
	\cline{4-5}
	 &  &  &  $\kappa_{\mu}$ & 1.002 \\
	\hline
	 &  &  &  $\kappa_t$ & 0.999  \\
	\hline
	 &  &  & $\kappa_{\gamma}$ & 1.077 \\
	\hline\hline
	\multicolumn{5}{|c|}{ $\chi^2_{\text{min}} = $ 4.14~~~~~~~~~~~~~~~~~ ${\chi^2_{\text{min}}}/$\texttt{dof} = 0.46~~~~~~~~~~~~~~~~\texttt{p-value} = 0.90} \\
	\hline\hline
		\end{tabular} 
		\caption{Displayed here are the best-fit and mean values, along with the 1$\sigma$ uncertainty for all the parameters and best-fit values for the observables in the NH scenario. The last row represents $\chi^2_{\text{min}}$, ${\chi^2_{\text{min}}}/$\texttt{dof} and \texttt{p-value}.}
	\label{tab:nh_input_output}
	\end{center}
\end{table}
From our MCMC analysis, the minimum $\chi^2$ value obtained for this scenario is $\chi^2_{\text{min}} = 4.14$, leading to $\chi^2_{\text{min}}/\texttt{dof} = 0.46$ and \texttt{p-value} = 0.90. The values of each parameter and observable corresponding to this best-fit point are detailed in Table \ref{tab:nh_input_output}. The masses of the three neutrinos at the best-fit point are $m_{\nu_1} = 2.42 \times 10^{-6}$ eV, $m_{\nu_2} = 8.65 \times 10^{-3}$ eV, and $m_{\nu_3} = 5.04 \times 10^{-2}$ eV, which automatically satisfies the limit on the sum of neutrino masses as mentioned in Table \ref{tab:neutrino_obs}. Only the third neutrino mass gets mass at the leading order as defined in Equation.~\ref{eq:third_neutrino_mass}. 
The choice of neutrino mass hierarchy is expected to be reflected in the resulting best-fit values and the $1\sigma$, $2\sigma$ and $3\sigma$ ranges of the $\lambda$ and $\lambda^{\prime}$ couplings. In NH of neutrino masses, the third mass eigenstate is considerably heavier compared to the other two states, and hence the best-fit value of $\lambda^{\prime}_{333}$ is almost twice that of $\lambda^{\prime}_{i33}$ ($i=1,2$) with $\lambda^{\prime}_{233}$ slightly larger than $\lambda^{\prime}_{133}$ as expected. The same trend is visible in best-fit values of $\lambda_{i33}$ ($i=1,2$) as well. 
These results are reflected in the best-fit values listed in the Table.~\ref{tab:nh_input_output}. 
\begin{figure}[!htb]
    \includegraphics[width=0.49\textwidth]{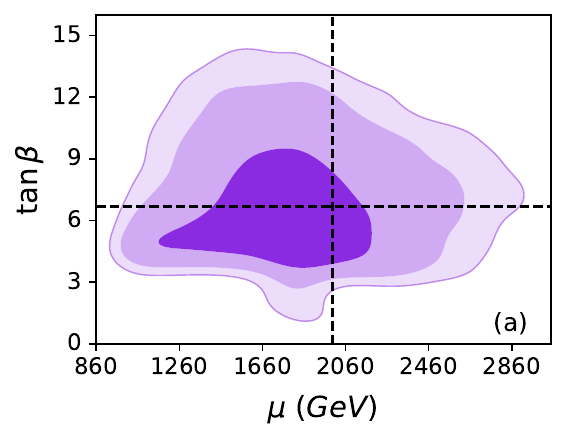}
    \includegraphics[width=0.49\textwidth]{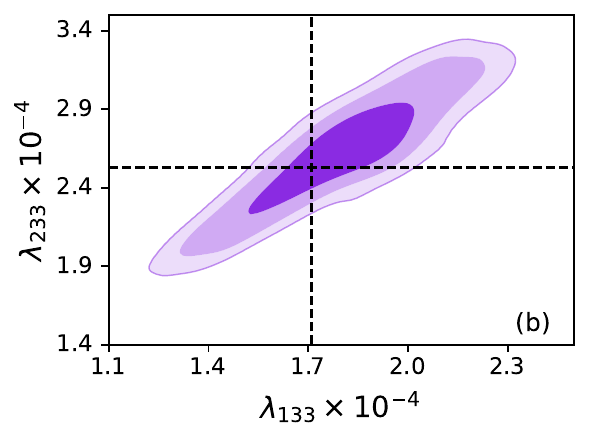}
    \includegraphics[width=0.49\textwidth]{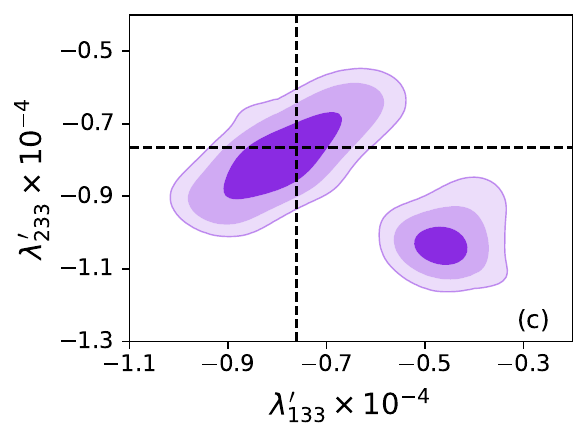}
    \includegraphics[width=0.49\textwidth]{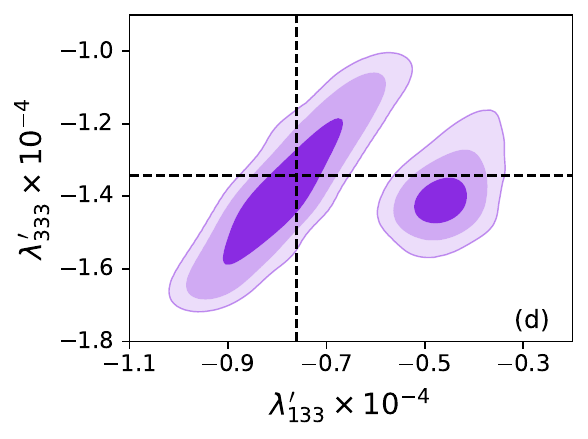}
    \caption{Marginalized 2-dimensional posterior distribution of the parameters in the NH scenario are shown on different planes: (a) $\mu-\tan\beta$ (b) $\lambda_{133}-\lambda_{233}$, (c) $\lambda_{133}^{\prime}-\lambda_{233}^{\prime}$, (d) $\lambda_{133}^{\prime}-\lambda_{333}^{\prime}$. The dark, lighter, and lightest violet colored regions represent contours at 68\%, 95\%, and 99\% C.L.
    The dashed lines represent the best-fit values of each parameter.}
    \label{fig:nh1}
\end{figure}
The mean values, along with $2\sigma$ allowed regions for each parameter, are also listed in Table.~\ref{tab:nh_input_output}. 

The 2-dimensional marginalized posterior distributions corresponding to different parameter planes are shown in Figure.~\ref{fig:nh1} where the dark, lighter, and lightest violet colors represent the 1$\sigma$, 2$\sigma$ and 3$\sigma$ regions, respectively. From these 2D contour plots, we have obtained that the allowed regions for $\mu$-$\tan\beta$ parameter space at 95\% (68\%) confidence level (C.L.) are 980-2640 (1160-2195) GeV and 2.70-12.70 (3.70-9.60) for $\mu$ and $\tan\beta$ respectively. Also the obtained parameter ranges for $\lambda_{133}$ and $\lambda_{233}$ parameters at 95\% C.L. are (1.31-2.32)$\times10^{-4}$  and (1.94-3.23)$\times10^{-4}$ respectively. Similarly the 2$\sigma$ allowed regions corresponding to \textit{LQD} type RPV couplings are -9.72$\times10^{-5}$ to -3.75$\times10^{-5}$, -1.13$\times10^{-4}$ to -0.59$\times10^{-4}$ and -1.67$\times10^{-4}$ to -1.05$\times10^{-4}$ for $\lambda_{133}^{\prime}$, $\lambda_{233}^{\prime}$ and $\lambda_{333}^{\prime}$ respectively. It is evident from both figures (c) and (d) in Fig.~\ref{fig:nh1} that the $\lambda_{i33}^{\prime}$ couplings parameters have more than one allowed region. 

The distribution of $\lambda_{i33}^{\prime}$ with $i = 1, 2$ exhibits a multimodal posterior, indicating that the model allows multiple solutions for these two parameters. This behavior primarily arises from the dependence on the third neutrino mass eigenstate, as described in Eq.~\ref{eq:third_neutrino_mass} which indicates that the mass crucially depends on $\sum_i \lambda_{i33}^2$ where, $i=1,2,3$. Once $\lambda_{333}^{\prime}$ picks up a relatively large value owing to normal hierarchy, the other two parameters can combine in more than one way to reproduce the same $\sum_i \lambda_{i33}^2$ factor.
The marginalized distribution of each parameter and the correlation of each parameter with every other parameter are shown in Fig.~\ref{fig:nh_corner} of Appendix~\ref{sec:appendix1}.

\subsubsection{Inverted Hierarchy scenario}
\label{sec:ih}
\begin{table}[!htb]
\begin{center}
	\begin{tabular}{||c|c|c||c|c||}
	\hline\hline
	 \multicolumn{3}{||c||}{Input parameters} & \multicolumn{2}{c||}{Best-fit observable}	\\
	\hline
     Parameter & Best-fit point & Mean value$^{+1\sigma}_{-1\sigma}$   & Observable & Best-fit \\  
	\hline\hline
	\multirow{2}{*}{$\lambda_{133}\times10^{-4}$} & \multirow{2}{*}{2.21} & \multirow{2}{*}{2.34$^{+0.21}_{-0.16}$} & $\Delta m_{21}^2$[eV$^2$] & 7.50$\times10^{-5}$  \\
	\cline{4-5}
	& &  & $\Delta m_{31}^2$[eV$^2$] & 2.54$\times10^{-3}$ \\
	\hline
	\multirow{2}{*}{$\lambda_{233}\times10^{-4}$} & \multirow{2}{*}{4.88} & \multirow{2}{*}{4.73$^{+0.11}_{-0.19}$} & $\theta_{13}$ & 8.54$^{\circ}$  \\
	\cline{4-5}
	& & & $\theta_{12}$ & 34.20$^{\circ}$ \\
	\hline
	\multirow{2}{*}{$\lambda_{133}^{\prime}\times10^{-4}$} & \multirow{2}{*}{1.29} & \multirow{2}{*}{1.28$^{+0.18}_{-0.08}$}  & $\theta_{23}$ & 49.31$^{\circ}$ \\
	\cline{4-5}
	 &  &  & $m_h$ [GeV] & 125.39\\
	\hline
	\multirow{2}{*}{$\lambda_{233}^{\prime}\times10^{-5}$} & \multirow{2}{*}{-5.92} & \multirow{2}{*}{-6.43$^{+0.54}_{-0.97}$} & $\mathcal{B}r(B \rightarrow X_s \gamma)$ & 3.14$\times10^{-4}$  \\
	\cline{4-5} 
	& &  & $\mathcal{B}r(B_s \rightarrow \mu^+ \mu^-)$ & 3.17$\times10^{-9}$  \\
	\hline
	\multirow{2}{*}{$\lambda_{333}^{\prime}\times10^{-4}$} & \multirow{2}{*}{-1.25} & \multirow{2}{*}{-1.22$^{+0.60}_{-0.06}$} & $\mathcal{B}r(B \rightarrow \tau \nu)$ & 1.25$\times10^{-4}$  \\
	\cline{4-5}
	 &  &  & $\kappa_z$ & 1.0  \\
	\hline
	\multirow{2}{*}{$\mu$ [GeV]} & \multirow{2}{*}{1676} & \multirow{2}{*}{1640$^{+30}_{-38}$}  & $\kappa_w$ & 1.0  \\
	\cline{4-5}
	& & & $\kappa_b$ & 1.001 \\
	\hline
	\multirow{2}{*}{$\tan\beta$} & \multirow{2}{*}{8.38} & \multirow{2}{*}{8.67$^{+0.53}_{-0.75}$}  & $\kappa_{\tau}$ & 1.001 \\
	\cline{4-5}
	 &  &  &  $\kappa_{\mu}$ & 1.001 \\
	\hline
	 &  &  &  $\kappa_t$ &  0.999  \\
	\hline
	 &  &  & $\kappa_{\gamma}$ & 1.098 \\
	\hline\hline
	\multicolumn{5}{|c|}{ $\chi^2_{\text{min}} = $ 4.56~~~~~~~~~~~~~~~~~ ${\chi^2_{\text{min}}}/$\texttt{dof} = 0.51~~~~~~~~~~~~~~~~\texttt{p-value} = 0.87} \\
	\hline\hline
		\end{tabular} 
		\caption{Displayed here are the best-fit and mean values, along with the 1$\sigma$ uncertainty for all the parameters and best-fit values for the observables in the IH scenario. The last row represents $\chi^2_{\text{min}}$ , ${\chi^2_{\text{min}}}/$\texttt{dof} and \texttt{p-value}.}
	\label{tab:ih_input_output}
	\end{center}
\end{table}

In this scenario, the lightest state ($\nu_3$) is dominantly tau-flavored. Both $\nu_1$ and $\nu_2$ are heavier and maintain the same hierarchy between them as in the NH scenario. The best-fit point, along with the $1\sigma$ uncertainty of each parameter, are shown in the Table.~\ref{tab:ih_input_output}. We have obtained the minimum $\chi^2$ as 4.56 corresponding to \texttt{dof} 9 which leads to $\chi^2_{\text{min}}/\texttt{dof}$ = 0.51 and \texttt{p-vale} = 0.87. The observable values corresponding to the best-fit point are also displayed in the Table.~\ref{tab:ih_input_output}. The neutrino masses at the best-fit point is $m_{\nu_{1}} = 5.05\times10^{-2}$ eV, $m_{\nu_{2}} = 5.12\times10^{-2}$ eV and $m_{\nu_{3}} = 3.58\times10^{-6}$ eV and the sum of these three neutrino masses are 0.1 which means it satisfies the cosmological limit mentioned in the Table.~\ref{tab:neutrino_obs}. 
\begin{figure}[!htb]
    \includegraphics[width=0.49\textwidth]{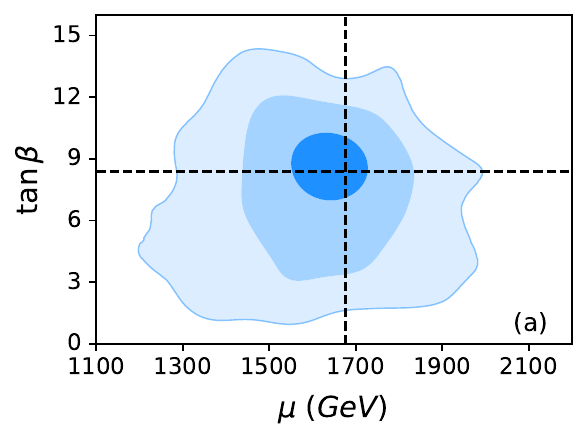}
    \includegraphics[width=0.49\textwidth]{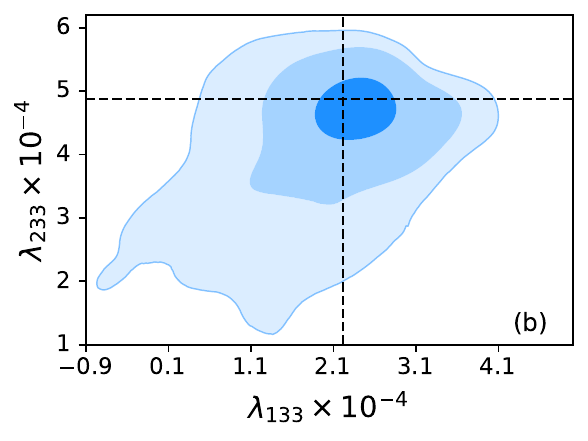}
    \includegraphics[width=0.49\textwidth]{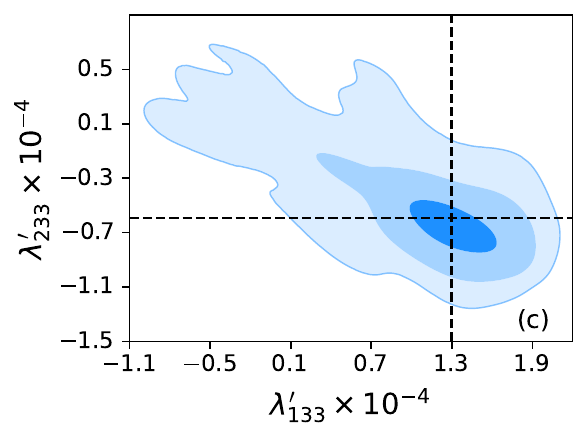}
    \includegraphics[width=0.49\textwidth]{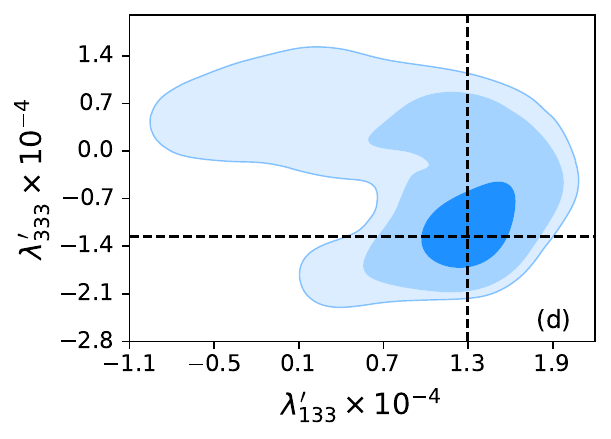}
    \caption{Marginalized 2-dimensional posterior distribution of the parameters in the IH scenario are shown on the different planes: (a) $\mu-\tan\beta$ (b) $\lambda_{133}-\lambda_{233}$, (c) $\lambda_{133}^{\prime}-\lambda_{233}^{\prime}$, (d) $\lambda_{133}^{\prime}-\lambda_{333}^{\prime}$. The dark, lighter, and lightest blue colored regions represent contours at 68\%, 95\%, and 99\% C.L.
    The dashed lines represent the best-fit values of each parameter.}
    \label{fig:ih1}
\end{figure}
As in the previous scenario, here also the main contribution to the dominantly tau-flavored neutrino mass eigenstate arises from $\lambda_{333}^{\prime}$ coupling with small admixtures arising from the other two nonzero $\lambda^{\prime}$ couplings. The other two light neutrino mass eigenstates get contribution from both $\lambda_{i33}$ and $\lambda_{i33}^{\prime}$ couplings, where $i$=1,2. Small mixing of these states with the dominantly tau-flavored state arises from $\lambda_{333}^{\prime}$ coupling as well. 

Figure.~\ref{fig:ih1} represents the marginalized distributions of the parameter space. Here, the dark and lighter and lightest blue colors refer to the $1\sigma$ and $2\sigma$ and $3\sigma$ regions, respectively, for each figure. The 2$\sigma$ allowed regions for $\mu$ and $\tan\beta$ obtained from the Fig.~\ref{fig:ih1} (a) are 1437-1826 GeV and 3.14-12.05 respectively. For \textit{LLE} type RPV couplings the allowed 2$\sigma$ regions are (1.24-3.66)$\times10^{-4}$ GeV and (3.52-5.50)$\times10^{-4}$ GeV corresponding to $\lambda_{133}$ and $\lambda_{233}$ respectively. Similarly for $\lambda_{133}^{\prime}$, $\lambda_{233}^{\prime}$ and $\lambda_{333}^{\prime}$ couplings the regions are (0.27-1.86)$\times10^{-4}$ GeV, -1.01$\times10^{-4}$ to -0.07$\times10^{-4}$ GeV and -1.97$\times10^{-4}$ to -0.87$\times10^{-4}$ GeV respectively. The marginalized distribution of each parameter and the correlation of each parameter with every other parameter are shown in Fig.~\ref{fig:ih_corner} of Appendix~\ref{sec:appendix1}.

\subsubsection{Possible Collider signals to probe allowed region}
It is worth discussing the possible collider signatures of the favored parameter space obtained with a bino LSP. The LSP mass here is chosen to be around 300 GeV, and the wino type lightest chargino ($\chonepm$) or the second lightest neutralino ($\lsptwo$) becomes the NLSP with a mass around $\sim$ 1.25 TeV. We revisit different possible leptonic final states arising from the wino pair production and the sensitivity of the multi-lepton LHC analyses to probe such scenarios. For the illustration purpose, we present the decay signatures of the LSP at the best-fit point in Table~\ref{tab:nh_br} for the NH scenario, while the input parameters are already mentioned in  Table.~\ref{tab:nh_input_output}. Note that these results correspond to the best-fit point. Due to the presence of only nonzero 
$\lambda_{i33}$ (i=1,2) and $\lambda_{j33}^{\prime}$ (j=1,2,3) coupling, the LSP decays to $\tau\tau\nu$ and $\tau l\nu_{\tau}$ final states with 51\% and 32\% branching ratios respectively via LLE couplings and the most dominant decay modes via LQD couplings are $bb\nu_{e/\mu/\tau}$ with a total 16\% branching fraction\footnote{For more detailed discussion on the possible decay channels and LHC signatures with LLE and LQD couplings see Refs.~\cite{Dreiner:2023bvs,Choudhury:2023eje}.}. 
 
\begin{table}[!htb]
\begin{center}
	\begin{tabular}{||c||c||}
	\hline\hline
	Final state from single LSP & Final state  from LSP pair \\
	\hline \hline
& $4\tau + \met$ = 26\%  \\ 
    	$\tau\tau\nu \equiv \tau\tau\nu_e (35.2\%)+ \tau\tau\nu_{\mu} (16.0\%)$    $\to$ 51\%    & $1l3\tau+\met$ = 33\%\\
    $\tau l\nu_{\tau} \equiv$  $\tau e\nu_{\tau} (3.6\%)+ \tau\mu\nu_{\tau} (28.5\%) \to$  32\% & $2l2\tau+\met$ = 10\% \\\cline{2-2} 
   & $2\tau2b+\met$ = 16\% \\\cline{1-1}
& $1\tau1l2b+\met$ = 10\% \\\cline{2-2}
     $bb\nu \equiv$  $bb\nu_e (3.1\%)+ bb\nu_{\mu} (3.1\%) + bb\nu_{\tau} (9.5\%) \to $  16\% & $4b+\met$ = 2.5\% \\
    $ltb \equiv etb  + \mu tb$ = 0.5\% & $1l2\tau1b1t+\met$ = 1\%\\ 
    $\tau tb$ = 0.7\% & Others = 1.5\%\\
	\hline\hline
		\end{tabular} 
		\caption{Branching ratios of different final sates at the best-fit point coming from single LSP  and the LSP pair are shown for the NH scenario.  Here $l$ denotes $e,\mu$ only. }
		
	\label{tab:nh_br}
	\end{center}
\end{table}

In Sec.~\ref{sec:constraints_mass}, we have already discussed the limits on wino masses coming from  $\chonepm \lsptwo$ production via $N_l \ge 4l$ and $3l1\tau$ final states, which are sensitive for $\lambda_{121/122}$  and $\lambda_{133/233}$ type LLE couplings. For scenarios with 
 single nonzero $\lambda_{i33}$ coupling, the LSP pair from the NLSP ($\chonepm, \lsptwo$) decays gives rise to $2l2\tau+\met$, $1l3\tau+\met$ and $4\tau+\met$ final states with 25\%, 50\% and 25\% branching ratios respectively and the limits on NLSP masses becomes 500 GeV weaker compared to scenarios with non zero $\lambda_{121/122}$ couplings. For tau enriched final states, the $3l1\tau$ final state search provides a stringent limit 
  on $\mchonepm/\mlsptwo$ ($\sim$ 1.025 TeV for a LSP with mass 300 GeV)\cite{ATLAS:2021yyr}. From our analysis, for the best-fit point, we observe that the total pure leptonic (including $\tau$)  branching fraction is about $\sim 83\%$, and this reduction will further weaken the mass limits. It may be noted that for our analysis, the NLSP mass ($\mchonepm/\mlsptwo$) is $\sim$ 1.25 TeV, which is allowed by current LHC data. In a recent analysis ~\cite{Choudhury:2023eje}, the authors have studied the future prospect of electroweakino searches for different choices of LLE couplings from multilepton final states. Furthermore, with $b$-jet enriched final states, one can probe the scenarios where both LLE and LQD couplings are nonzero, but a dedicated collider analysis in this regard is beyond the scope of this paper. 
 However, using the 
ATLAS $N_l \ge 4l$ analysis \cite{ATLAS:2021yyr}, validated in Refs.\cite{Choudhury:2023eje, Choudhury:2023yfg}, we observe that the lower bounds on $\mchonepm$ 
reduce to $\sim$ 900 GeV when both the  LLE and LQD couplings are nonzero, and as a result, the relevant branching ratios get suppressed. 
  
\begin{figure}[!h]
\begin{center}
    \includegraphics[width=0.7\textwidth]{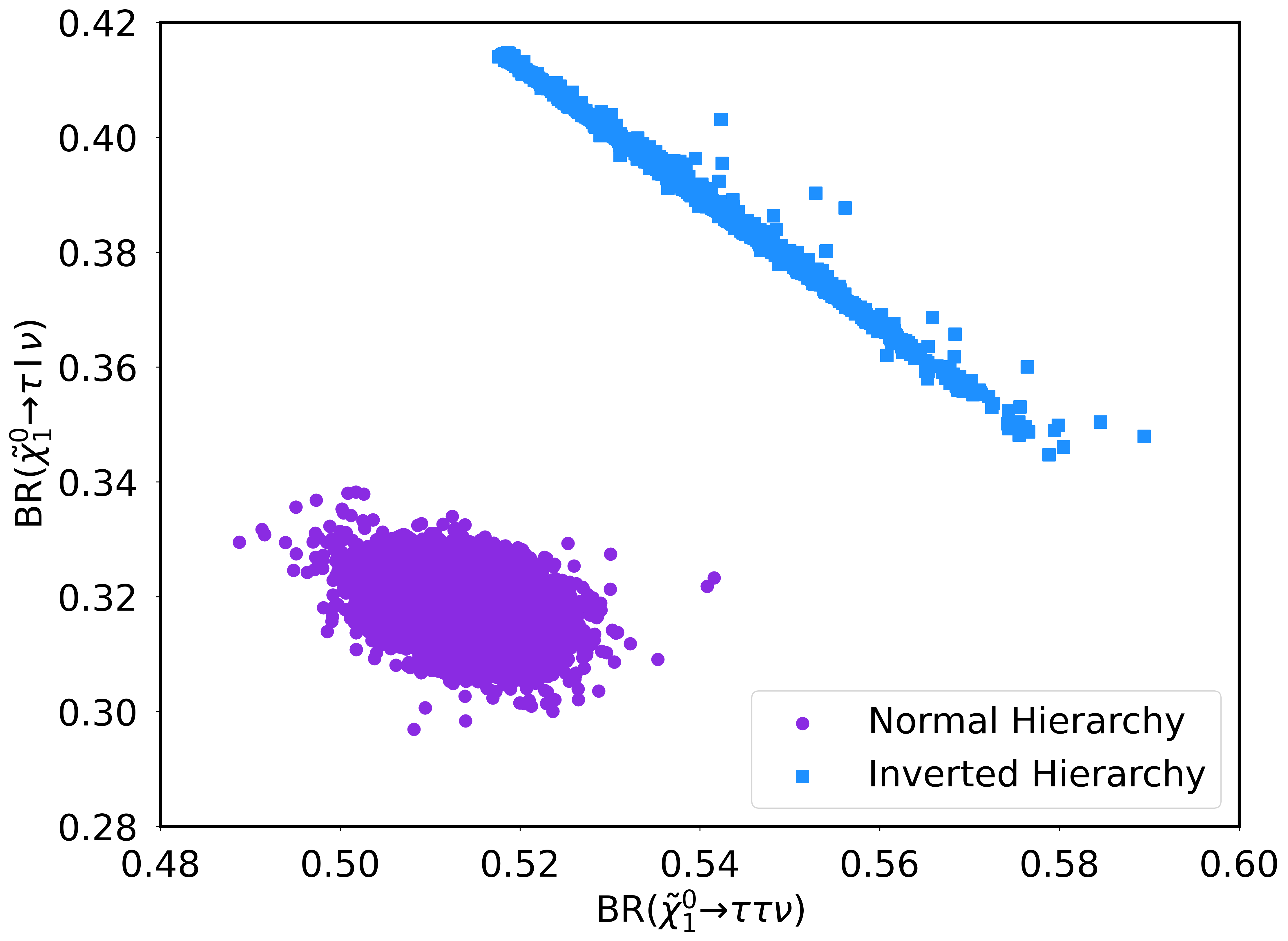}
    \caption{$\mathcal{B}r(\lspone \to \tau\tau\nu$) and $\mathcal{B}r(\lspone \to \tau l\nu$) are shown for NH and IH scenarios.}
    \label{fig:br_compare}
    \end{center}
\end{figure}
Compared to the NH scenario, the branching ratios of the decays driven by the LLE couplings increase by $\sim$ 10\% in the IH scenario. This is a direct effect of the resultant hierarchy obtained in the values of the trilinear couplings in our best-fit point. We analyze randomly chosen 10\% points within the $3\sigma$ allowed region and present the variations in branching ratios for the final states $\tau\tau\nu$ and $\tau l\nu$ in Fig.~\ref{fig:br_compare}. From the figure, it is evident that the points corresponding to different hierarchy scenarios are clustered in two separate regions. The branching ratios corresponding to IH regions are higher than those for NH points. Consequently, at the LHC, final states with a higher lepton multiplicity are expected in the IH scenario.
\begin{table}[!htb]
\centering
\resizebox{\textwidth}{!}{%
\renewcommand{\arraystretch}{1.1}
	\begin{tabular}{|c||c|c|c|c|c|c|c||c|c|}
	\hline
	Bench-  & \multicolumn{7}{c||}{Input parameters} & \multicolumn{2}{c|}{Branching}\\
	\cline{2-8} 
	mark & $\lambda_{133}$ & $\lambda_{233}$ & $\lambda_{133}^{\prime}$ & $\lambda_{233}^{\prime}$ & $\lambda_{333}^{\prime}$ & \multirow{2}{*}{$\mu$ (GeV)} & \multirow{2}{*}{$\tan\beta$} & \multicolumn{2}{c|}{ratio}  \\ \cline{9-10}
	points& $\times 10^{-4}$ & $\times 10^{-4}$ & $\times 10^{-4}$ & $\times 10^{-4}$ & $\times 10^{-4}$ & & & $\tau \tau \nu$ & $\tau l \nu$ \\
	\hline \hline
	BP$_{\text{NH}}$ & $1.96$ & $2.46$ & $-0.50$ &  $-0.98$ & $-1.42$ & 1824.61 & 6.36 & 0.524 & 0.306 \\
	\hline
	BP$_{\text{IH}}$ & $2.32$ & $4.97$ & $1.35$ &  $-0.63$ & $-1.23$ & 1630.16 & 8.12 & 0.521 & 0.414 \\
	\hline\hline
	\multicolumn{4}{|c}{Masses for both benchmark points} & \multicolumn{6}{|c|}{Signal Yield at $\sqrt{s} = 14$~TeV \& $\mathcal{L} = 3000$~fb$^{-1}$} \\
	\hline
	\multicolumn{4}{|c}{\multirow{2}{*}{$m_{\chonepm/\lsptwo} = 1000$~GeV, $m_{\lspone} = 300$~GeV}} &  \multicolumn{3}{|c|}{BP$_{\text{NH}}$} & \multicolumn{3}{c|}{BP$_{\text{IH}}$} \\
	\cline{5-10}
		\multicolumn{4}{|c}{ }& \multicolumn{3}{|c}{82} & \multicolumn{3}{|c|}{108} \\
	\hline
		\end{tabular} 
}
		\caption{The input parameters and the branching ratio corresponding to $\tau\tau\nu$ and $\tau l \nu$ final states are shown for two benchmark points corresponding to two hierarchies are displayed. Here $l$ denotes $e,\mu$ only. BP$_{\text{NH}}$ and BP$_{\text{IH}}$ refers to the benchmark points for Normal and Inverted hierarchy respectively.}		
	\label{tab:nh_ih_br}%
\end{table}
To illustrate this further, we select two benchmark points from two different scenarios with masses $m_{\chonepm/\lsptwo} = 1000$~GeV and $m_{\lspone} = 300$~GeV as shown in Table~\ref{tab:nh_ih_br}. We evaluate the number of signal events at the High-Luminosity LHC (HL-LHC) with $\sqrt{s} = 14$~TeV and $\mathcal{L} = 3000~\mathrm{fb}^{-1}$ for these benchmark points. For that, we consider the electroweakino pair productions: $pp \rightarrow \chonepm\lsptwo + \chonepm\chonemp$, where $\chonepm \rightarrow \lspone W^{\pm}$ ($W^{\pm} \rightarrow l^{\pm} \nu$) and $\lsptwo \rightarrow \lspone Z$ ($Z \rightarrow l^{\pm}l^{\mp}$), with $\lspone$ decaying via both $LLE$ and $LQD$ type couplings. As mentioned earlier, the ATLAS Collaboration has already looked for these electroweakino pair production for with $N_l \geq 4$ final state using Run-II data~\cite{ATLAS:2021yyr}. Using the same search with $N_l \geq 4$, we estimate that $\sim$ 82 and 108 signal events can be obtained at the HL-LHC for the NH and IH scenarios, respectively, assuming that object reconstruction efficiencies and other detector-related factors remain unchanged. The estimated signal yields indicate that there is $\sim$ 30\% enhancement for IH scenario as compared to NH scenario. As a result, a careful measurement of the final state events in the above-mentioned leptonic channels can provide an alternate probe of neutrino mass hierarchy at the LHC.  
\section{RPV Model with a stop LSP (\texttt{Stop$_{\lambda \lambda^\prime}$-model})}
\label{sec:squark_lsp}
In this section, we analyze the \texttt{Stop$_{\lambda \lambda^\prime}$-model} where the lighter top squark or stop ($\st1$) is the LSP. The mixing effect of the weak eigenstates of the stop in the stop mass matrix is expected to be large due to the presence of the top mass in the off-diagonal term  $m_t(A_t - \mu / tan\beta)$. Thus, the lighter stop mass ($\mst1$) may become much smaller than all other squarks and sparticles\footnote{It may be noted that the renormalization group running from the GUT scale to the weak scale may also suppress the stop mass~\cite{Martin:1997ns}.}. This model-independent large mixing feature could give rise to a theoretically well-motivated stop LSP scenario, as well as a stop NLSP scenario\footnote{Several phenomenological groups have analyzed light stop scenarios in the context of both RPC and RPV models~\cite{Bandyopadhyay:2010ms, Choudhury:2012kn, Chakraborty:2015xia, Konar:2016ata, Han:2016xet, Chamoun:2014eda, Chakraborty:2015wga, Das:2005mr, Datta:2009dc, Bose:2014vea, Datta:2000yc, Restrepo:2001me, Marshall:2014kea, Marshall:2014cwa}.}.

\begin{table}[!htb]
\begin{center}
	\begin{tabular}{||c|c|c||c|c||}
	\hline\hline
	 \multicolumn{3}{||c||}{Input parameters} & \multicolumn{2}{c||}{Best-fit observable}	\\
	\hline
     Parameter & Best-fit point & Mean value$^{+1\sigma}_{-1\sigma}$   & Observable & Best-fit \\  
	\hline\hline
	\multirow{2}{*}{$\lambda_{133}\times10^{-4}$} & \multirow{2}{*}{1.56} & \multirow{2}{*}{1.64$^{+0.10}_{-0.10}$} & $\Delta m_{21}^2$[eV$^2$] & 7.50$\times10^{-5}$  \\
	\cline{4-5}
	& &  & $\Delta m_{31}^2$[eV$^2$] & 2.55$\times10^{-3}$ \\
	\hline
	\multirow{2}{*}{$\lambda_{233}\times10^{-4}$} & \multirow{2}{*}{2.29} & \multirow{2}{*}{$2.39^{+0.11}_{-0.11}$} & $\theta_{13}$ & 8.52$^{\circ}$  \\
	\cline{4-5}
	& & & $\theta_{12}$ & 34.30$^{\circ}$ \\
	\hline
	\multirow{2}{*}{$\lambda_{133}^{\prime}\times10^{-5}$} & \multirow{2}{*}{-5.55} & \multirow{2}{*}{5.56$^{+0.24}_{-0.35}$}  & $\theta_{23}$ & 49.06$^{\circ}$ \\
	\cline{4-5}
	 &  &  & $m_h$ [GeV] & 125.39\\
	\hline
	\multirow{2}{*}{$\lambda_{233}^{\prime}\times10^{-5}$} & \multirow{2}{*}{-5.54} & \multirow{2}{*}{-6.43$^{+0.54}_{-0.97}$} & $\mathcal{B}r(B \rightarrow X_s \gamma)$ & 3.30$\times10^{-4}$  \\
	\cline{4-5} 
	& &  & $\mathcal{B}r(B_s \rightarrow \mu^+ \mu^-)$ & 3.17$\times10^{-9}$  \\
	\hline
	\multirow{2}{*}{$\lambda_{333}^{\prime}\times10^{-4}$} & \multirow{2}{*}{-0.98} & \multirow{2}{*}{-1.01$^{+0.42}_{-0.51}$} & $\mathcal{B}r(B \rightarrow \tau \nu)$ & 1.24$\times10^{-4}$  \\
	\cline{4-5}
	 &  &  & $\kappa_z$ & 1.0  \\
	\hline
	\multirow{2}{*}{$\tilde{m}_{q_{3L}}$[GeV]} & \multirow{2}{*}{1499} & \multirow{2}{*}{1505$^{+20}_{-19}$}  & $\kappa_w$ & 1.0  \\
	\cline{4-5}
	& & & $\kappa_b$ & 1.001 \\
	\hline
	\multirow{2}{*}{$\tan\beta$} & \multirow{2}{*}{7.03} & \multirow{2}{*}{6.08$^{+0.53}_{-0.59}$}  & $\kappa_{\tau}$ & 1.001 \\
	\cline{4-5}
	 &  &  &  $\kappa_{\mu}$ & 1.001 \\
	\hline
	 &  &  &  $\kappa_t$ &  0.999  \\
	\hline
	 &  &  & $\kappa_{\gamma}$ & 1.098 \\
	\hline\hline
	\multicolumn{5}{|c|}{ $\chi^2_{\text{min}} = $ 2.93~~~~~~~~~~~~~~~~~ ${\chi^2_{\text{min}}}/$\texttt{dof} = 2.93/9 = 0.32~~~~~~~~~~~~~\texttt{p-value} = 0.96} \\
	\hline\hline
		\end{tabular} 
		\caption{The best-fit and mean values with the 1$\sigma$ uncertainty 
		for the seven free parameters are shown for stop-LSP model in the NH scenario. Last two columns represent best-fit values of the observables. The last row represents $\chi^2_{\text{min}}$, ${\chi^2_{\text{min}}}/$\texttt{dof} and \texttt{p-value}.}
	\label{tab:nh_stop}
	\end{center}
\end{table} 

For the MCMC analysis, the parameters are varied/fixed as mentioned in the last column of Table.~\ref{tab:parameter_range}. We vary the masses of third generation left-handed squarks in the range 1-3 TeV, and to obtain the $\st1$ as the LSP, we choose bino, wino, Higgsino masses as $M_1 = M_2 = \mu = m_{\tilde{q}{3L}} + 1000$ GeV, all three generation slepton masses as $m_{\tilde{l}} = m_{\tilde{q}_3L} + 500$ GeV and all other squarks, 
gluino ($M_3$) masses and Higgs mass parameter ($M_A$) fixed at 3 TeV. To achieve a Higgs mass around 125 GeV, we fix $A_t$ at 4.5 TeV. For the  \texttt{Stop$_{\lambda \lambda^\prime}$-model}, we only focus on the normal hierarchy scenario with seven free parameters. The best-fit point and the $1\sigma$ uncertainty of each parameter, along with the observables values at the best-fit point, are shown in Table.~\ref{tab:nh_stop}. We found that the MCMC analysis leads to $\chi^2_{\text{min}}$ as 2.93, and  \texttt{p-value} becomes 0.96. This scenario provides a better fit to the observable $\mathcal{B}r(B \rightarrow X_s \gamma$) owing to the favored top and bottom squark masses obtained here and thus improves the overall fit quality. The values and hierarchies of the $\lambda_{i33}$ and $\lambda_{i33}^{\prime}$ follow a similar pattern with the NH scenario of the \texttt{Bino$_{ \lambda \lambda^\prime}$-model} as discussed in Sec.~\ref{sec:both_coupling}, but the allowed parameter space is more constrained for this model. The corner plot for this stop-LSP model is shown in Fig.~\ref{fig:stop_nh_corner} (see Appendix~\ref{sec:appendix3}). Now, we proceed to analyze the characteristics of stop-LSP scenarios in the context of LHC.

\subsection{Possible Collider Signatures at the LHC and impacts of Run-II data}
In general, to satisfy the neutrino oscillation data, the RPV couplings are required to be small. In spite of these tiny couplings, $\st1$ decays via 
RPV modes to a bottom quark and a charged lepton with 100\% branching ratio in the stop-LSP model \cite{Datta:2000yc,Restrepo:2001me,Bose:2014vea,Das:2005mr,Datta:2009dc,Marshall:2014kea,Marshall:2014cwa}. Hence the final states from stop pair production will consist of 
an opposite sign dilepton pair along with two hard \texttt{b-tagged} jets ($p p \to \st1 \st1^* \to (b \ell^+) (\bar b \ell^-)$, where $\ell$ denotes $e, \mu, \tau$). The ATLAS Collaboration has recently conducted a search for stop pair production using the 13 TeV full Run-II dataset and obtained lower limits on $\mst1$ for different choices of branching ratios of three leptonic decay modes~\cite{ATLAS:2024zkx}. It is important to note that the relative branching ratios among the three leptonic decay modes depend on the choices of the hierarchies of LQD couplings and thus can be useful to discriminate the 
 neutrino mass hierarchy in case the LHC experiment observes any $2b2\ell$ signal. Additionally, we aim to adjudge the effect of the latest LHC limits~\cite{ATLAS:2024zkx} on the allowed mass ranges of $\st1$, which also satisfy the neutrino oscillation results.
\begin{figure}[!htb]
\begin{center}
    \includegraphics[width=0.9\textwidth]{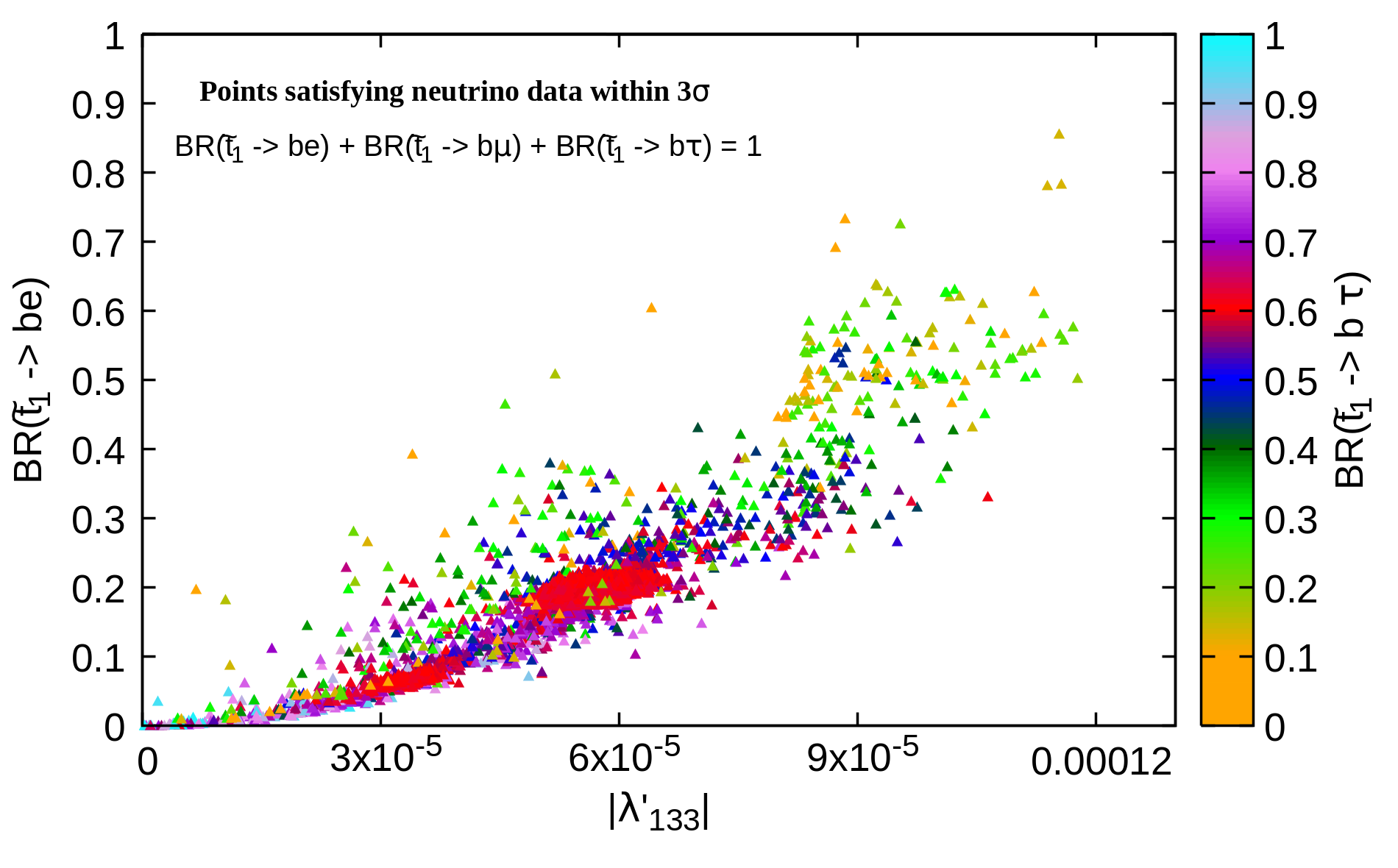}
    \caption{$\mathcal{B}r(\st1 \to be$) as a function of $|\lambda_{133}^{\prime}|$. The $\mathcal{B}r(\st1 \to b\tau$) branching ratio is also represented by the color palette.}
    \label{fig:br_lambdap133}
    \end{center}
\end{figure}
As mentioned earlier,  the most relevant constraints on the allowed parameter space come from the light stop pair ($\st1 \st1^*$) production, and the limits depend on the relative branching ratios among $\st1 \to be, b\mu,$ and $ b\tau$, which crucially depends on the hierarchies of $\lambda_{i33}^{\prime}$ couplings. As mentioned, even with tiny  $\lambda_{i33}^{\prime}$ couplings, $\st1$ decays to $b\ell~(\ell =e,\mu$, and $\tau)$ with 100\% branching fractions instead of RPC decays and indirect decays via LLE couplings. It is expected that the branching ratios of the decay modes $\st1 \to b\ell$ will depend on the corresponding $\lambda_{i33}^{\prime}$ coupling, which is reflected in Fig.~\ref{fig:br_lambdap133}. 

\begin{figure}[!htb]
\begin{center}
    \includegraphics[width=0.9\textwidth]{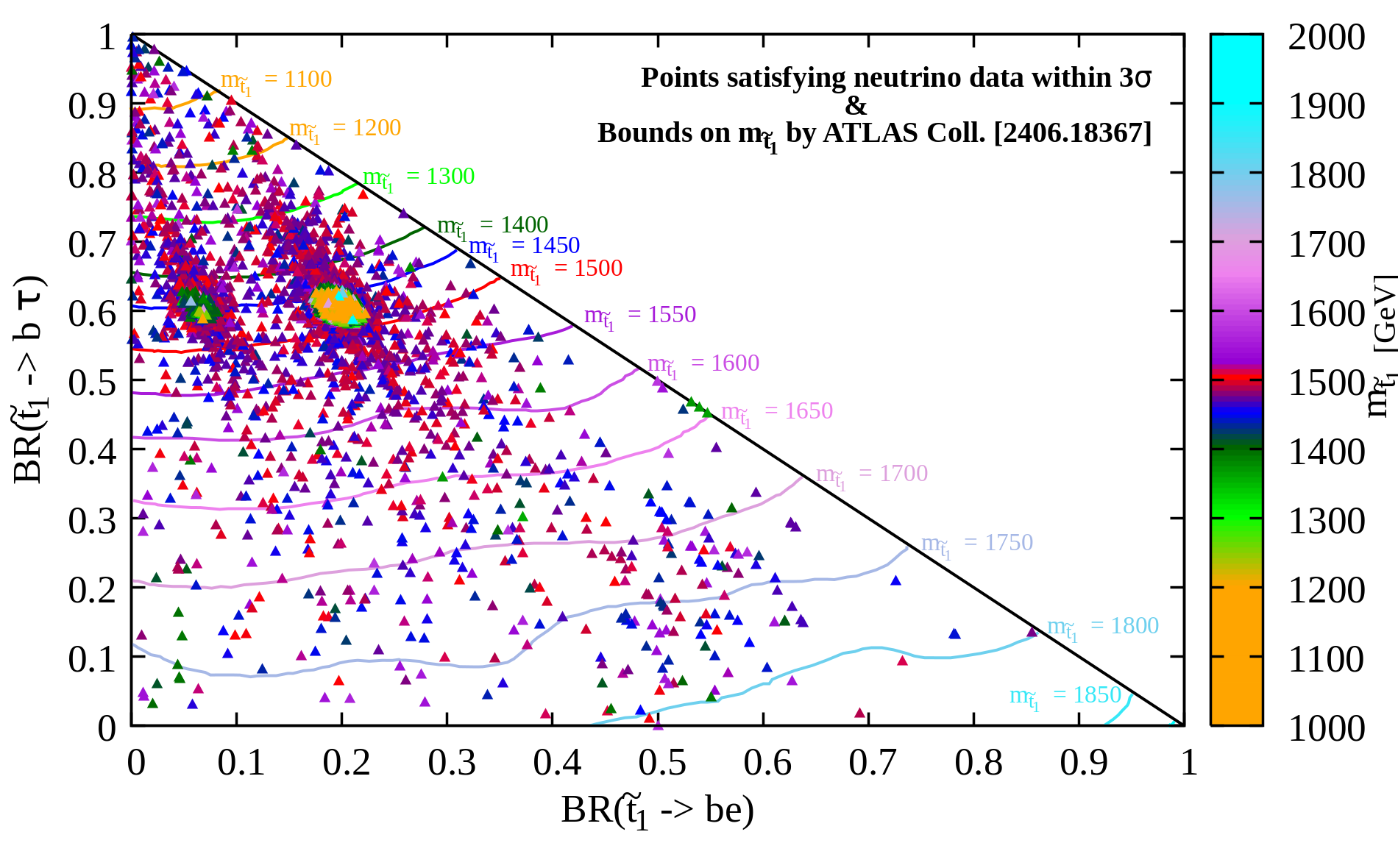}
        \includegraphics[width=0.49\textwidth]{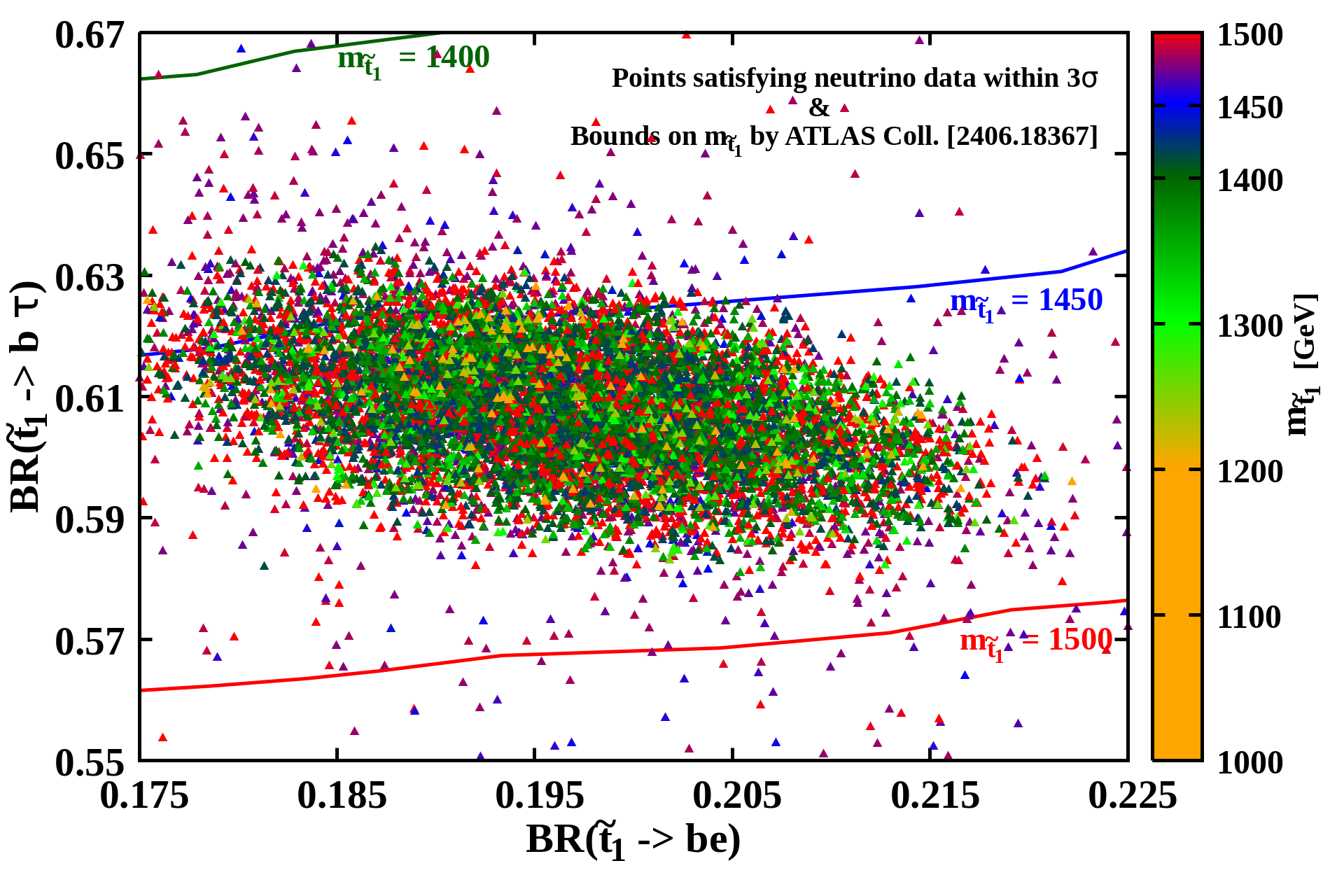}
    \includegraphics[width=0.49\textwidth]{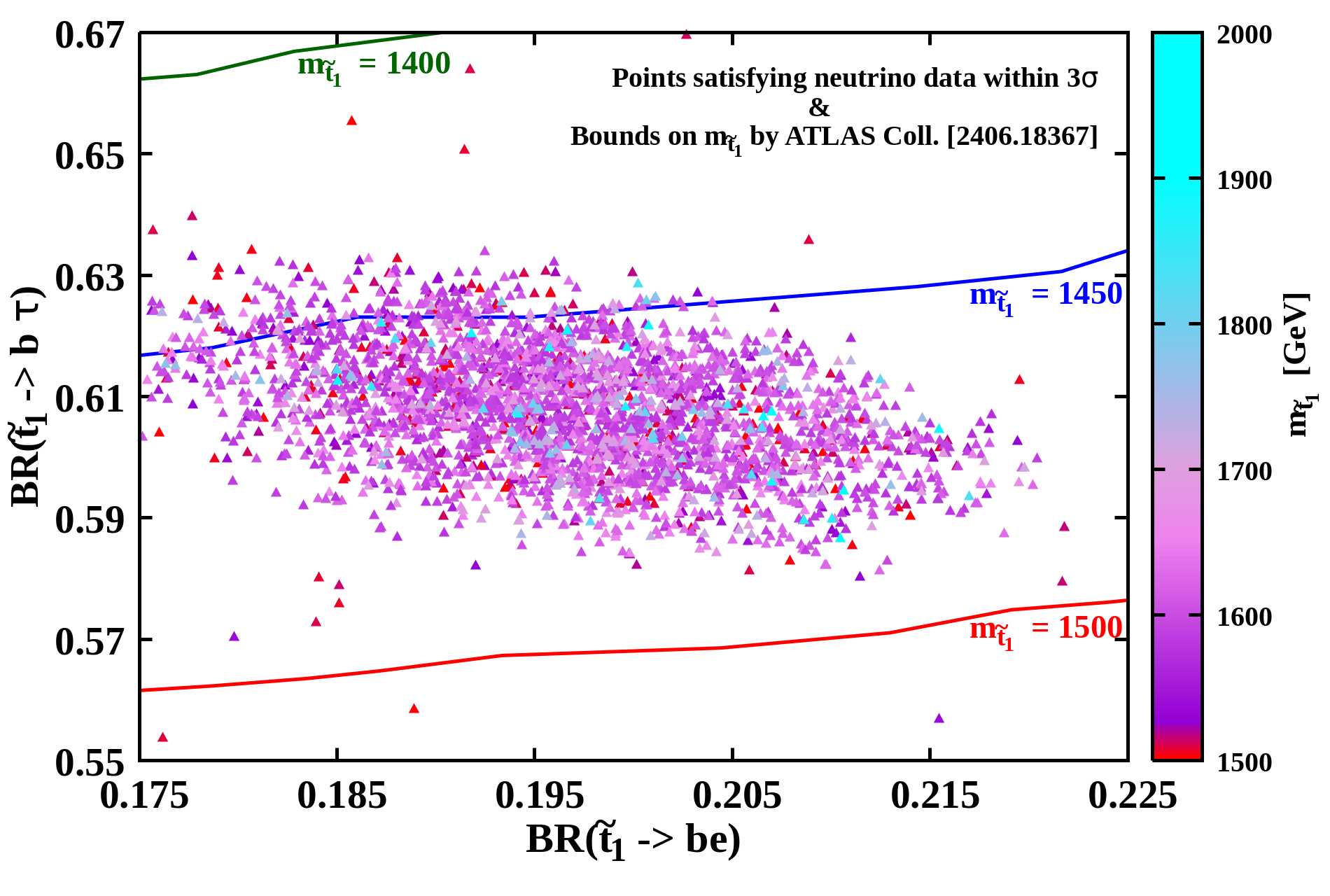}
    \caption{The interplay of LHC bounds~\cite{ATLAS:2024zkx} on $\mst1$ as a function of the $\st1$  branching ratios shown, where total $\mathcal{B}r(\st1 \to be+ b\mu + b\tau$) is unity. The different colored lines represent the highest $\mst1$ excluded by the ATLAS Collaboration at 95\% C.L. subjected to choice of branching ratio~\cite{ATLAS:2024zkx}. The points allowed from the MCMC analysis at $3\sigma$ level are shown in the top panel, and the variation of $\mst1$ of these points is illustrated by the color palette with a similar color convention as the limit contours. Thus, any parameter space point with a particular color, which lies below (above) the similar colored exclusion contour, is excluded (allowed) by the LHC Run-II data. Zoomed-in distribution around the best-fit region presented with  $\mst1 < 1500$ GeV in the bottom left panel and $\mst1 \ge 1500$ GeV in the bottom right panel.  }
\label{fig:stop_br_full_all}
\end{center}
\end{figure}

Here we choose the points\footnote{From our final data-set, we have randomly chosen a small fraction ($\sim 6\%$) of the points to present the results in 
Fig.\ref{fig:br_lambdap133} and Fig.\ref{fig:stop_br_full_all}.} with $\chi^2 \le \chi^2_{\text{min}} + 11.83$, which corresponds to the allowed $\chi^2$ value for 2D parameter space at $3\sigma$ level \cite{ParticleDataGroup:2024cfk}. For illustration purpose we present the $\mathcal{B}r(\st1 \to be$) as a function of  $\lambda_{133}^{\prime}$ and also show the $\mathcal{B}r$($\st1 \to b \tau$) through the color palette  in Fig.~\ref{fig:br_lambdap133}. 
 $\mathcal{B}r(\st1 \to be$) gradually increases with higher values  $\lambda_{133}^{\prime}$ and reaches upto $\sim $ 85\%, while $\mathcal{B}r(\st1 \to b \tau$) decreases as $\mathcal{B}r(\st1 \to be+ b\mu + b\tau$) is always unity.  The dense red region corresponds to the adjacent region of our best-fit point (with $\mst1$ = 1470 GeV), where $\mathcal{B}r(\st1 \to be/b\mu) \sim$ 0.2 and  $\mathcal{B}r(\st1 \to b \tau)\sim$ 0.6. For our best fit point, large $b\tau$ branching arises due to the fact that   the   $\lambda_{333}^{\prime}$ coupling is larger than  $\lambda_{133/233}^{\prime}$ by a factor of two (see Table~\ref{tab:nh_stop}). The LHC sensitivity reduces with the increase of $b\tau$ branching ratio. For example, the lower limit  on $\mst1$ reduces from 1900 GeV to 800 GeV  in scenarios with a 100\% branching ratio to $b e$ compared to those with a 100\% branching ratio to 
  $b \tau$~\cite{ATLAS:2024zkx}. We observe that the branching ratio for the $b\tau$ mode can vary from 0 to 100\% in our MCMC-allowed parameter region. 
  
In Fig.\ref{fig:stop_br_full_all}, we illustrate the impact of existing exclusion limits on $\mst1$ on our allowed parameter space. For this purpose, we have used results obtained by the ATLAS Collaboration~\cite{ATLAS:2024zkx} and present them in the $\mathcal{B}r(\st1 \to be)$ vs $\mathcal{B}r(\st1 \to b \tau$) plane. The various colored lines represent the highest limits on $\mst1$ at a 95\% C.L. for a specific point on the branching ratio plane while the total leptonic ($\st1 \to be+ b\mu + b\tau$) branching ratio is assumed to be unity. We use a similar color convention for the representation of the $\mst1$ of 3$\sigma$ allowed points through the color palette. For example, the scattered blue and red points correspond to the parameter space where $\mst1$ is around 1450 and 1500 GeV, respectively, and the LHC data excludes those points with $\mathcal{B}r(\st1 \to b \tau$) is $\lesssim$ 0.6. Parameter space within the mass range   $1350 < \mst1 < 1400$ GeV (dark green points) are still beyond the reach of current ATLAS analysis where $\mathcal{B}r(\st1 \to b \tau )>$ 0.7. The most densely populated region around the best-fit point,  with $\mathcal{B}r(\st1 \to be) \sim$ 0.2 and  $\mathcal{B}r(\st1 \to b \tau)\sim$ 0.6, includes parameters space in the stop mass range of $1 <\mst1 < 2$ TeV. However, this feature is not clearly visible due to the overlap. Therefore, we zoom in on this region for better illustration purposes and present the results for two mass ranges $\mst1 < 1500$ GeV (bottom left panel) and $\mst1 \ge 1500$ GeV (bottom right panel) in the Fig.~\ref{fig:stop_br_full_all}. These zoomed plots show that parameter space with light stop mass up to 1350 GeV is excluded while points with $\mst1 > 1600$ GeV escape the limits. Note that while generating these plots, we have sampled the allowed region around the best-fit point more extensively, which explains the sparseness of the points in other regions. The key features of these plots are as follows: 
 \begin{itemize}
 \item All the yellow points ($1000<\mst1<1200$ GeV) and the light green points ($\mst1 < 1350$ GeV) lie below the corresponding exclusion line. Even a branching ratio value of $\mathcal{B}r(\st1 \to be + b\mu) \sim $ 0.4 is sufficient to exclude these points. 
 \item All dark/light cyan and magenta points correspond to $\mst1 > 1600$ GeV 
 and $\mathcal{B}r(\st1 \to b \tau) \sim $ 0.6, and these are allowed by the current LHC data. 
 \item The dark green, blue, and red points, corresponding to stop masses around 1400, 1450, and 1500 GeV, respectively, are excluded at 95\% C.L. if the branching ratio of $b\tau$ mode is $<$ 50\%. To be precise, a $b\tau$ branching ratio of 72\%, 70\%, and 65\% for these three masses makes the parameter space allowed by the current LHC data. These allowed parameter regions are worth exploring in the context of the future HL-LHC experiment\footnote{Note that, we have not explored the IH scenario with stop LSP. One can expect a comparatively smaller branching ratio of the stop LSP in the $b\tau$ channel in this scenario, leading to more stringent constraints.}. 
 \end{itemize}

\section{Conclusion}
\label{sec:conclusion}
$R$-parity violating supersymmetric theories allow us to include lepton and baryon number violating interaction terms in the Lagrangian. This has immense phenomenological implications. In this work, we have explored scenarios with specific lepton number violating interactions in the context of the latest experimental results. Our main objective has been to ascertain the allowed ranges of these unknown RPV couplings and other relevant SUSY parameters so that they can fit the existing neutrino oscillation data precisely. In addition, we have also considered the SM-like Higgs precision data and B-meson decay branching ratios as observables since their prediction may also be affected while fitting neutrino oscillation data. While performing the analysis, we ensured that the whole parameter space considered in this study also respected the existing collider constraints. We have considered two different scenarios characterized by the choice of the LSP, namely, bino and stop. The analysis has been performed with the help of an MCMC algorithm that not only identifies the best-fit point but can also be used to generate the allowed regions of the model parameter space through likelihood estimation. 

We have presented our results in the form of multiple two-dimensional marginalized posterior distributions in different parameter planes, where different confidence level contours are clearly shown. Different trilinear RPV scenarios in the context of neutrino oscillation have been explored before, but this is the first attempt that goes beyond just choosing some benchmark points. What makes our results robust and unique is the fact that we not only provide the whole allowed region, but our results also provide the likelihood of different parts of that allowed region. That gives us an idea of which regions are likely to be probed in the near future. In addition to that, limits on trilinear couplings mostly exist in the literature in product forms, which allows one of the couplings in the product to be somewhat unrestricted compared to the other. Our results put robust constraints on the trilinear couplings individually. Furthermore, any future study on these models must adhere to these constrained parameter spaces presented in this article. Our study indicates that the trilinear couplings $\lambda_{i33}$ ($i$=1,2) and $\lambda_{j33}^{\prime}$ ($j$=1,2,3) are all confined to the $10^{-4}-10^{-5}$ range, whereas, $\tan\beta$ is restricted to values lower than 15. After identifying the allowed regions, we proceeded to discuss what sort of collider signal one can expect for the two different LSP scenarios and comment on their search status in light of the existing collider results. We have observed that the allowed regions obtained from our analysis for the bino-LSP scenarios are mostly allowed by the LHC electroweakino searches while the stop pair production searches partially exclude the parameter space obtained for the stop-LSP scenario. The relevant decay branching ratios of the LSPs for these direct searches reflect the choice of hierarchy and the abundance of the corresponding final state events can be an indirect probe of the neutrino mass hierarchy.


\section{Acknowledgement}
\label{sec:ack}
A.C. and S.M. acknowledge the Anusandhan National Research Foundation, India, for supporting this work under Core Research Grant No. CRG/2023/008570. S.M. further acknowledges additional support from the Anusandhan National Research Foundation under Core Research Grant No. CRG/2022/003208.    

\newpage

\appendix
\section{Corner plot for the model with bino LSP}
\label{sec:appendix1}
\begin{figure}[!htb]
    \includegraphics[width=1\textwidth]{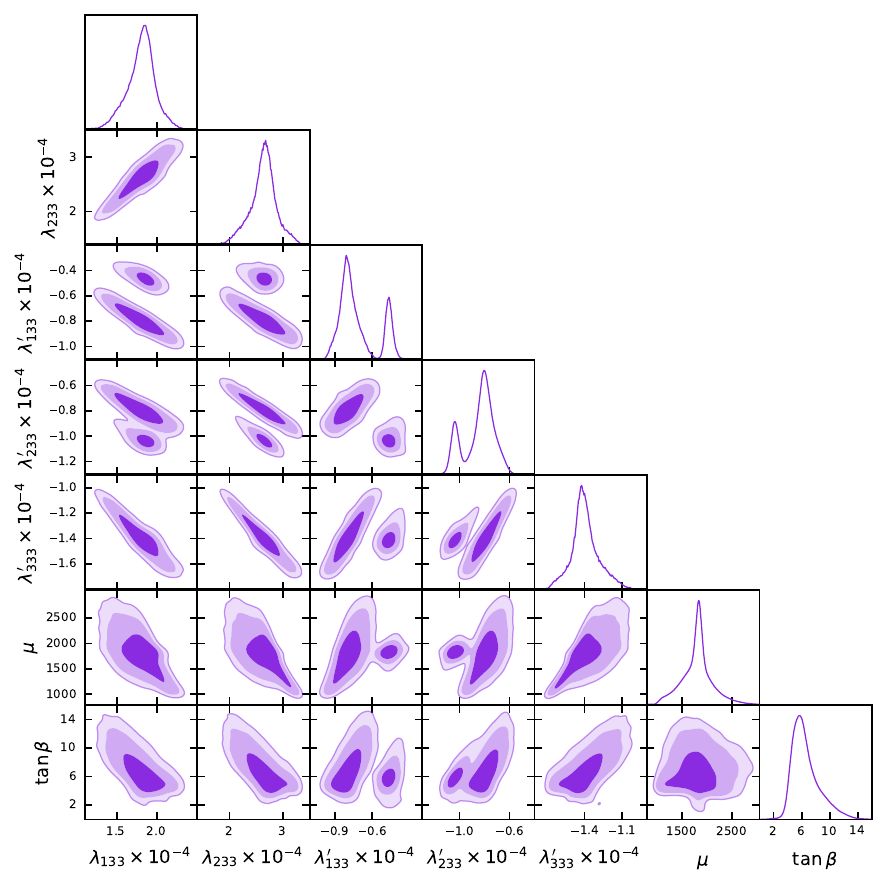}
    \caption{Corner plot for the input parameters in NH scenario. The diagonal histograms are 1D posterior probability distributions, and the other contour plot show the covariances between parameters. The 1$\sigma$, 2$\sigma$, and 3$\sigma$ regions are represented by different shades of violet: dark violet, lighter violet, and lightest violet, respectively.}
    \label{fig:nh_corner}
\end{figure}

\newpage

\begin{figure}[!htb]
    \includegraphics[width=1.0\textwidth]{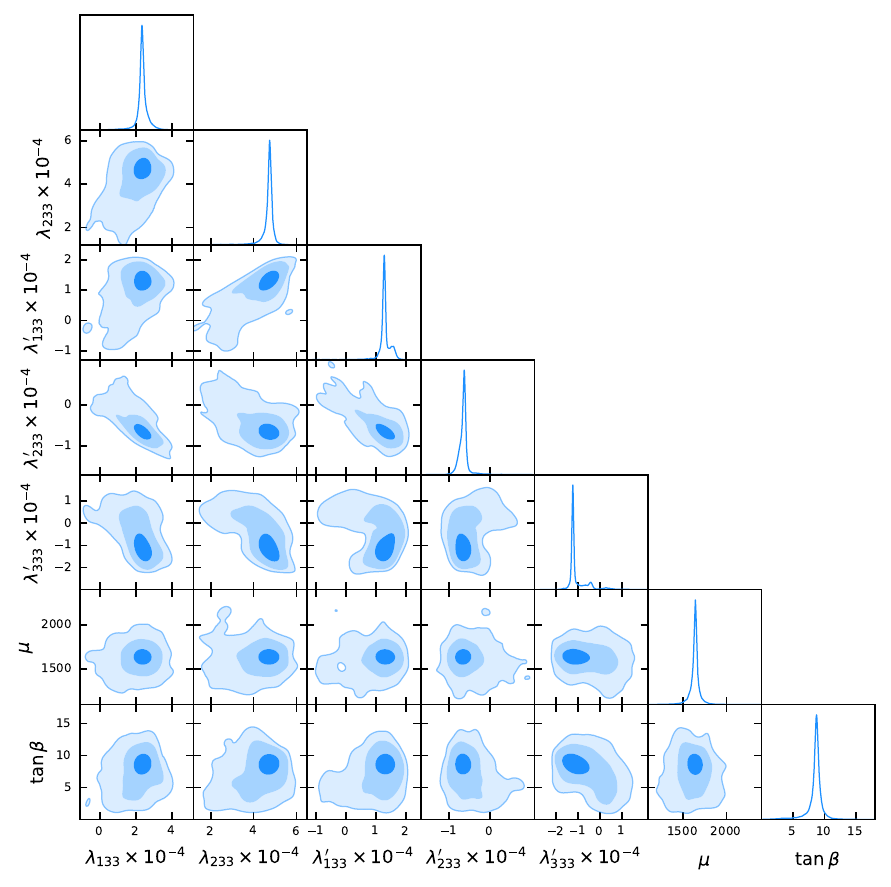}
    \caption{Corner plot for the input parameters in IH scenario. The diagonal histograms are 1D posterior probability distributions, and the other contour plot show the covariances between parameters. The 1$\sigma$, 2$\sigma$, and 3$\sigma$ regions are represented by different shades of blue: dark blue, lighter blue, and lightest blue, respectively.}
    \label{fig:ih_corner}
\end{figure}

\newpage

\section{Model with nonzero soft RPV coupling (\texttt{Bino$_{\lambda \lambda^\prime}^{A A^\prime}$-model})}
\label{sec:appendix2}
\label{se:soft}
For completeness, we also have considered a \texttt{Bino$_{\lambda \lambda^\prime}^{A A^\prime}$-model} where the soft coupling parameters corresponding to $\lambda_{i33}$ and $\lambda_{i33}^{\prime}$, represented as $A_{i33}$ and $A^{\prime}_{i33}$ respectively, are also included. These soft couplings introduce five additional parameters in the analysis, hereby increasing the model's complexity. For simplicity, we assume that all $A_{i33}$ have the same value, and similarly, all $A_{i33}^{\prime}$ have the same value. The range of these two parameters was set to [-5000, 5000] GeV. The rest of the parameters are the same as listed in the third column of Table.~\ref{tab:parameter_range}.
\begin{figure}[!htb]
\begin{center}
    \includegraphics[width=0.7\textwidth]{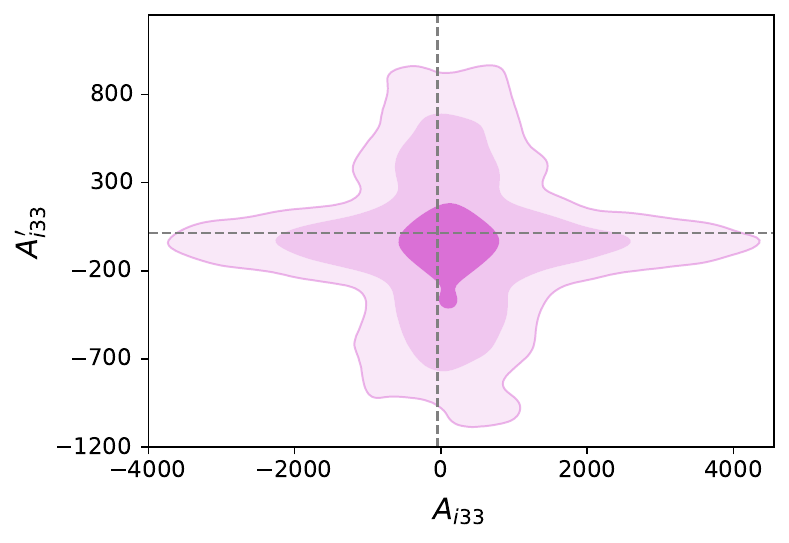}
    \caption{The posterior distribution of $A_{i33}$ and $A_{i33}^{\prime}$ is shown here. The dark, lighter, and lightest magenta colors represent contours at 68\%, 95\%, and 99\% C.L. Here, we have considered that two $A_{i33}$ have the same value and similarly, all the $A_{i33}^{\prime}$ have the same value. Light neutrino masses follow NH.}
    \label{fig:ap_app}
    \end{center}
\end{figure}
For this analysis, we assume a normal hierarchy in the light neutrino masses. There are 9 free parameters and 16 observables, resulting in 7 degrees of freedom. We have employed similar likelihood based MCMC analysis for \texttt{Bino$_{\lambda \lambda^\prime}^{A A^\prime}$-model} with NH scenario and obtain minimum $\chi^2$  as 4.13 ($\chi^2_{\text{min}}/\texttt{dof}$ = 0.59), which is similar to the \texttt{Bino$_{ \lambda \lambda^\prime}$-model} as discussed in the Sec.~\ref{sec:nh}. This is quite expected since the $A$ and $A^{\prime}$ couplings do not have any direct contributions to the light neutrino masses and mixings. The best-fit value, mean values, and $1\sigma$ uncertainty of the parameters, along with the values of observables at the best-fit point, are listed in Table~\ref{tab:nh_input_output_soft}. We have shown the 68\%, 95\%, and 99\% C.L. contours corresponding to these two parameters in the Figure.~\ref{fig:ap_app}. From this plot, we can see that $A_{i33}$ and $A_{i33}^{\prime}$ can have $2\sigma$ region as -2300 GeV to 2600 GeV and -710 to 775 GeV, respectively. Note that the values of $A_{i33}$ and $A_{i33}^{\prime}$ at the best-fit point are -45 GeV and 13 GeV, respectively, which are very close to the choice ($A_{i33}$ = $A_{i33}^{\prime}$ = 0) we made for other scenarios.   
\begin{table}[!htb]
\begin{center}
	\begin{tabular}{||c|c|c||c|c||}
	\hline\hline
	 \multicolumn{3}{||c||}{Input parameters} & \multicolumn{2}{c||}{Best-fit observable}	\\
	\hline
     Parameter & Best-fit point & Mean value$^{+1\sigma}_{-1\sigma}$   & Observable & Best-fit \\  
	\hline\hline
	\multirow{2}{*}{$\lambda_{133}\times10^{-4}$} & \multirow{2}{*}{1.82} & \multirow{2}{*}{1.82$^{+0.44}_{-1.03}$}  & $\Delta m_{21}^2$[eV$^2$] & 7.49$\times10^{-5}$  \\
	\cline{4-5}
	 &  &  & $\Delta m_{31}^2$[eV$^2$] & 2.55$\times10^{-3}$ \\
	\hline
	\multirow{2}{*}{$\lambda_{233}\times10^{-4}$} & \multirow{2}{*}{2.69} & \multirow{2}{*}{2.65$^{+0.52}_{-0.98}$}  & $\theta_{13}$ & 8.53$^{\circ}$  \\
	\cline{4-5}
	& & & $\theta_{12}$ & 34.06$^{\circ}$ \\
	\hline
	\multirow{2}{*}{$\lambda_{133}^{\prime}\times10^{-4}$} & \multirow{2}{*}{-0.45} & \multirow{2}{*}{-0.77$^{+1.25}_{-0.51}$}  & $\theta_{23}$ & 49.30$^{\circ}$ \\
	\cline{4-5}
	 &  &  & $m_h$ [GeV] & 124.88\\
	\hline
	\multirow{2}{*}{$\lambda_{233}^{\prime}\times10^{-4}$} & \multirow{2}{*}{-1.04} & \multirow{2}{*}{-0.81$^{+1.45}_{-0.38}$}  & $\mathcal{B}r(B \rightarrow X_s \gamma)$ & 3.15$\times10^{-4}$  \\
	\cline{4-5} 
	& & & $\mathcal{B}r(B_s \rightarrow \mu^+ \mu^-)$ & 3.18$\times10^{-9}$  \\
	\hline
	\multirow{2}{*}{$\lambda_{333}^{\prime}\times10^{-4}$} & \multirow{2}{*}{-1.41} & \multirow{2}{*}{-1.41$^{+0.92}_{-0.25}$}  & $\mathcal{B}r(B \rightarrow \tau \nu)$ & 1.24$\times10^{-4}$  \\
	\cline{4-5}
	 &  & & $\kappa_z$ & 1.0  \\
	\hline
	\multirow{2}{*}{$\mu$ [GeV]} & \multirow{2}{*}{1827} & \multirow{2}{*}{1851$^{+193}_{-226}$}  & $\kappa_w$ & 1.0  \\
	\cline{4-5}
	& &  & $\kappa_b$ & 1.002 \\
	\hline
	\multirow{2}{*}{$\tan\beta$} & \multirow{2}{*}{5.94} & \multirow{2}{*}{5.96$^{+3.22}_{-2.66}$}  & $\kappa_{\tau}$ & 1.002 \\
	\cline{4-5}
	 &  &  &  $\kappa_{\mu}$ & 1.002 \\
	\hline
	$A_{i33}$ & -45 & 105$^{+1183}_{-1198}$ &  $\kappa_t$ & 0.999  \\
	\hline
	$A_{i33}^{\prime}$ & 13 & -35$^{+445}_{-463}$ & $\kappa_{\gamma}$ & 1.077 \\
	\hline\hline
	\multicolumn{5}{|c|}{ $\chi^2_{\text{min}} = $ 4.13~~~~~~~~~~~~~~~~~ ${\chi^2_{\text{min}}}/$\texttt{dof} = 0.59~~~~~~~~~~~\texttt{p-value} = 0.75} \\
	\hline\hline
		\end{tabular} 
		\caption{The best-fit and mean values, along with the 1$\sigma$ uncertainty  for all the parameters and best-fit values for the observables in the NH scenario of the model with nonzero soft RPV couplings are shown here. The last row represents $\chi^2_{\text{min}}$, corresponding $\chi^2_{\text{min}}/\texttt{dof}$ and \texttt{p-value}.}
	\label{tab:nh_input_output_soft}
	\end{center}
\end{table}

\newpage

\section{Corner plot for the model with stop LSP}
\label{sec:appendix3}
\begin{figure}[!htb]
    \includegraphics[width=1.0\textwidth]{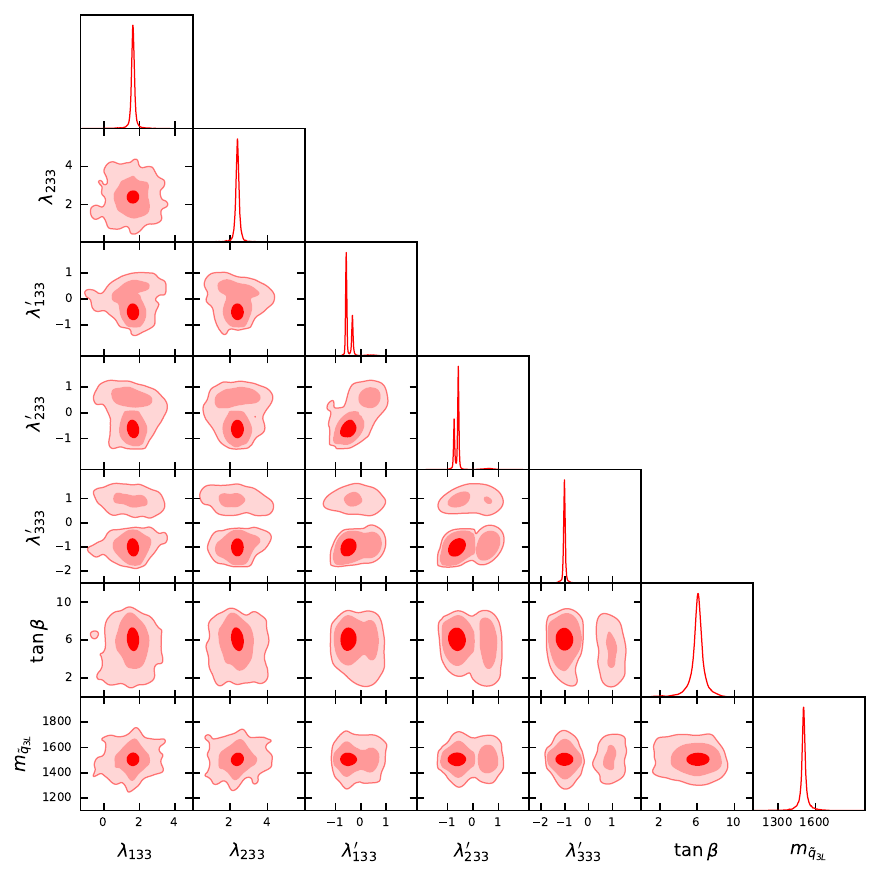}
    \caption{Corner plot for the input parameters in the NH scenario for the model with stop LSP. The diagonal histograms are 1D posterior probability distributions, and the other contour plot show the covariances between parameters. The 1$\sigma$, 2$\sigma$, and 3$\sigma$ regions are represented by different shades of red: dark red, lighter red, and lightest red, respectively.}
    \label{fig:stop_nh_corner}
\end{figure}

\newpage

\bibliography{reference}

\newpage

\end{document}